\documentstyle[11pt,newpasp,twoside,epsf]{article}
\def\edcomment#1{\iffalse\marginpar{\raggedright\sl#1\/}\else\relax\fi}
\reversemarginpar

\begin{document}
\title{Codes for optically thick and hot photoionized media -
 Radiative transfer and new developments}

\author{Anne-Marie Dumont}
\author{Suzy Collin}
\affil{Observatoire de Paris, Section de Meudon, Place Janssen, F-92195 Meudon, 
France}

\begin{abstract}
We describe a code designed for hot media {(T $\ge$ } a few 
10$^4$ K), optically thick to Compton scattering. It computes the 
structure of a plane-parallel slab of gas in thermal and 
ionization equilibrium, illuminated on one side or on both
sides by a given spectrum.  This code has been presented in a previous 
paper (Dumont, Abrassart \& Collin 2000), where several aspects were already 
discussed. So we focus here mainly on the recent 
developments. Presently the code solves the transfer of the continuum 
with the Accelerated Lambda Iteration method (ALI) and that of 
the lines in a two stream Eddington
approximation, without using the local escape probability formalism to 
approximate the line transfer. This transfer code is coupled with a
Monte Carlo code which allows to take  into account direct and 
inverse Compton diffusions, and to compute the spectrum emitted up to 
MeV energies, in any geometry. 
The influence of a few physical parameters is shown, and the importance of 
the density and pressure distribution (constant density, pressure equilibrium, or 
hydrostatic equilibrium) is stressed. Recent
improvements in the treatment of the atomic data are
 described, and foreseen developments are
 mentioned.
\end{abstract}

\section{Introduction}

 The average 
 continuum observed in Active Galactic Nuclei (AGN) is shown in 
 Fig. 1
from the optical to X-ray range, as obtained from a 
composite optical-UV spectrum of Francis et al. 
(1991) and Zheng et al. (1997), and from an average (slightly modified) soft X-ray
 spectrum of Laor et al.
(1997). It is widely admitted since several years that the optical-UV part 
of this continuum is produced by a warm (temperature of the 
 order of $10^5\ -\ 10^6$ K), optically thick or effectively thick medium, 
 most likely an accretion disc. Moreover, one deduces from the spectral 
 distribution in the hard X-ray range, in particular 
 the 30 keV hump observed in many Seyfert 1 galaxies, and from the 
 presence of other features like the iron K line 
around 7 keV, that this medium is irradiated by a hard X-ray source, which 
is partly reprocessed and reemitted (improperly named ``reflected"). So 
one sees a combination 
of the primary and of the reflected spectra (Ross \& Fabian 1993, and 
many subsequent works).  In another
 interpretation the UV-X continuum is produced by clouds from a disrupted 
 disk (Collin-Souffrin et al. 1996, Czerny \& Dumont 1998), and in this case one 
 would also see the spectrum emitted by the non illuminated surface of the 
 clouds. In both interpretations the 
 UV-soft X spectrum is emitted by a dense and thick shell (density higher 
 than  10$^{12}$ cm$^{-3}$, Thomson thickness higher than unity),  irradiated 
 by an X-ray continuum. Similar conditions are met in X-ray binary stars. 

\begin{figure}
\plotone{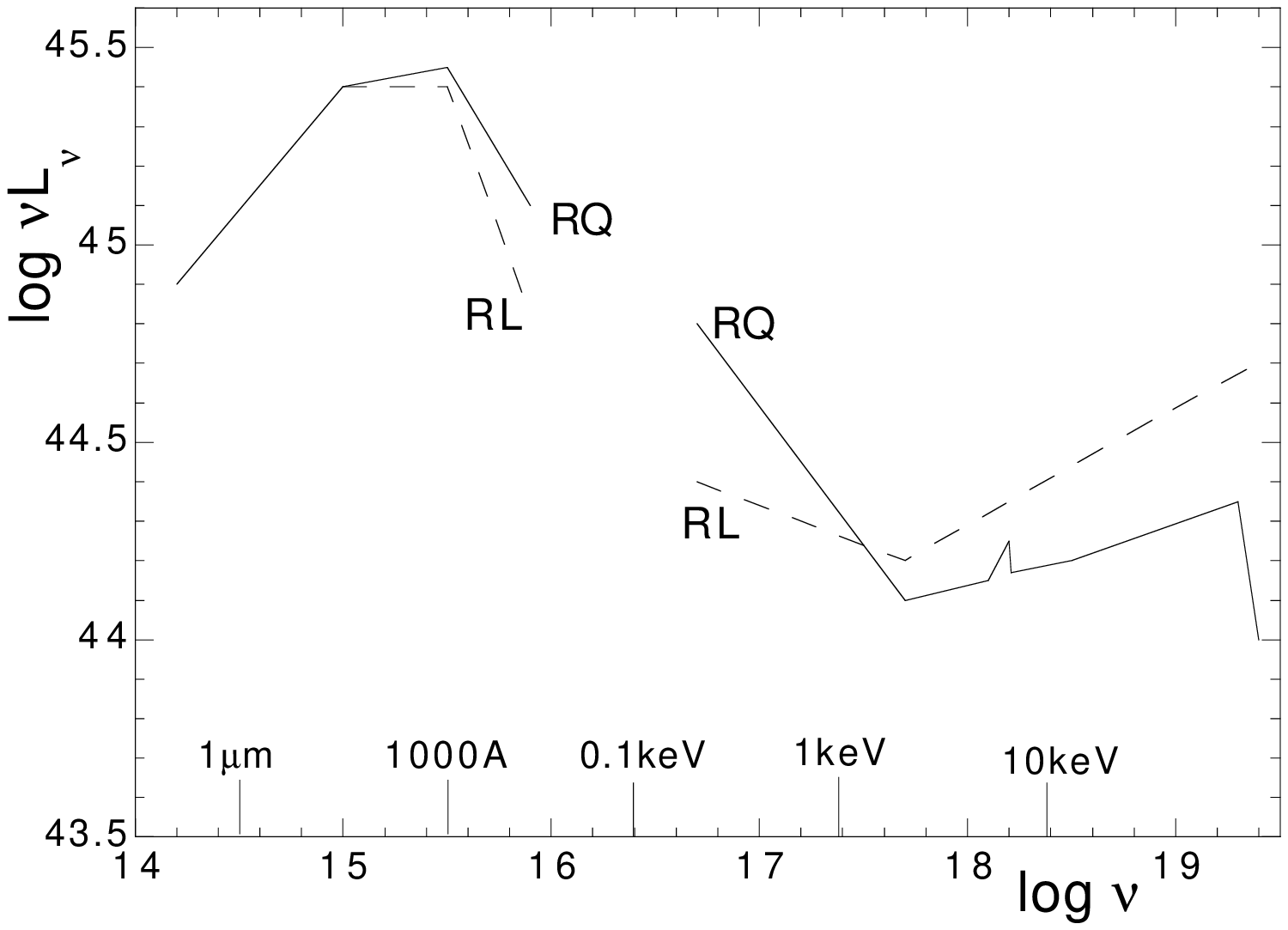}
\caption{Average continuum of radio loud and radio quiet 
Active Galactic Nuclei, in 
arbitrary units. }
\label{fig-AGNtypicbis}
\end{figure}

It is necessary to compute the structure of such an irradiated 
medium to predict the observed spectrum. Therefore one has to build 
``intermediate''codes, between those designed 
for planetary nebulae or Narrow or Broad Line Regions in AGN, and those 
designed for stellar atmospheres. These codes have to be valid for thick or 
semi-infinite dense media, eventually hot, irradiated by a non thermal 
continuum extending in the hard X-rays, to cover the whole range of 
situations encompassed in AGN and in binary stars.

Owing to the high optical thickness of the medium in several frequency 
ranges, such codes require that the transfer of both the continuum and 
the lines be solved in an ``exact'' way, that is avoiding approximations 
such as local escape approximation (for the lines) or one stream 
approximation (for the continuum). Since the medium is generally dense 
and sometimes close to Local Thermodynamical Equilibrium (LTE), they require 
that all processes and inverse processes be carefully handled. 
Being irradiated by an X-ray continuum, the medium contains a large number 
of ionic species, from low to high ionization, which should all be 
introduced in the computation. Finally, the medium being hot and thick, 
not only Thomson, but also Compton scattering, should be taken into account. 

We have undertaken to build a code which satisfies these requirements, 
 to compute the emission spectrum 
produced by irradiated thick and hot media like those commonly assumed in AGN 
close to the black hole, in a wide range of photon energy. Precisely we have built several interconnected 
codes,  which allow more 
flexibility. The ensemble is far from being perfect and still contains 
several approximations which restrict its use, but we improve it gradually.

TITAN is designed to study 
the structure of a warm or hot thick photoionized gas, and to compute its 
emission - reflection - transmission spectrum from the infrared up to about 20 
keV. It solves the energy balance, the ionization and the statistical 
equilibria, the transfer equations, in a plane-parallel geometry, for the lines 
and the continuum. Then, given the thermal and ionization stratification, the 
computation of the emitted spectrum from 1 keV to a {few hundreds} keV is 
performed with NOAR which uses a Monte-Carlo method taking into account 
direct and inverse Compton scattering (it allows also to study various 
geometries). In a previous paper (Dumont, Abrassart \& Collin 2000)  
TITAN and its interconnection with NOAR were described. We recall 
here the main characteristics of the code, and we describe recent 
improvements as well as new results, 
focussing only on the aspects which are not treated in a standard 
way. 

We briefly summarize below the physical processes (Section 2), the transfer 
method 
and the iteration procedure
(Section 3). The influence of the physical parameters, of the 
atomic data, and of the density 
distribution, are discussed in Section 4. The coupling of TITAN and NOAR is briefly 
described in Section 5.

\section{Generalities about TITAN}

\subsection{Physical processes}

In TITAN, the physical state of the gas (temperature, ionic abundances 
and level populations of all ionic species), is computed at each depth, 
assuming stationary state, i.e.: local balance between ionizations 
and recombinations, local balance between excitations and deexcitations, 
local energy balance, and finally total energy balance (equality 
between inward and outward fluxes).

The ionization equilibrium equations include 
radiative ionizations by continuum and line photons, collisional 
ionizations and recombinations, radiative and dielectronic recombinations, 
charge transfer by H and He atoms, the Auger effects, and ionizations by 
high energy electrons arising from ionizations by X-ray photons.
The emission-absorption mechanisms of the continuum include free-free 
and free-bound processes, two-photon process, and Thomson electron 
scattering.	
Inverse processes (except autoionization for dielectronic recombinations, 
and  recombinations onto excited levels for some ions, see next section) 
are computed through the equations of detailed balance. Induced processes 
are taken into account.
Energy balance equations include the same processes and Compton heating/cooling.

\subsection{Atomic data}

Hydrogen and hydrogen-like ions are treated as 5 or 6-level atoms. 
In a first step, in order to save computation time, non hydrogen-like ions 
were treated with a rough approximation: 
interlocking between 
excited levels was neglected and the populations of excited levels were
computed separately using a two-level approximation.  This approximation 
does not  predict correctly the details of the line spectrum, including 
the resonance lines themselves. Moreover ionizations 
from excited levels were not taken into account, though
recombinations  onto 
excited states were taken into account in the ionization equilibrium. This 
approach is 
is not correct close to LTE, and even far from LTE, if ones wants to 
predict accurately detailed spectral features (cf. Section 4.2). However it 
gives a correct overall spectral distribution in this last case. Such are the 
upper layers of an irradiated atmosphere where at the same time the 
fractional abundances of heavy elements in high ionization states are 
important, and the ground level of these ions is strongly overpopulated with respect to 
LTE. But it is not the case in the deep layers of a disk atmosphere, 
emitting the UV part of the spectrum, where 
the density is high, and the influence of the external irradiation is small.

This is why helium-like ions treated as complete 8-level atoms
and lithium-like ions treated as complete 5-level atoms have been recently
implemented in the code. All processes are thus taken into 
account for each level. This treatment is valid as far as the upper levels 
are in LTE with the continuum. To handle intermediate cases, we replace the 
missing upper levels which are very 
important especially for helium-like ions by adding an additional 
recombination rate, 
not compensated by ionization.

 In the future we plane to add several other levels, 
and to sum the contributions of the higher levels to handle at the same 
time
cases close to and cases far from LTE. 

Photoionization cross sections are fitted from Topbase (cf. Cunto et al
1993). Cross sections of neutral and once ionized heavy elements are not well 
represented, but they do not have any influence on these hot media. 

The gas composition include 10 elements (H, He, C, N, O, Ne, Mg, Si,
S, Fe), and all their ionic species.

The slab is divided in about 300 layers with variable thicknesses.
The code can work with a prescription for the density or for the pressure, 
such as a constant pressure or a pressure corresponding to 
hydrostatic equilibrium.

\subsection{Illumination}

The external illumination can be characterized by an ``ionization parameter", 
which we choose here as 
$\xi = {4\pi F_{\rm inc} / n_{\rm H}}$,
$F_{\rm inc}$ being the frequency integrated flux incident on one side 
of the slab and $n_{\rm H}$ the hydrogen number density at this surface. 
Note that there are several other definitions of the ionization parameter, 
for instance $U$, or $\Xi$,
but they correspond basically to the same concept, namely a
parameter which determines the ionization state. 
It is possible to take also into account a flux incident on the back side of
 the slab. 

Presently we mimic a semi-isotropic illumination, but with the ALI transfer 
method
which is implemented  (cf. below)
 it will be   
possible to take into account the angular distribution of the 
 incident radiation.

\section{Radiation transfer}

\begin{figure}
\plotone{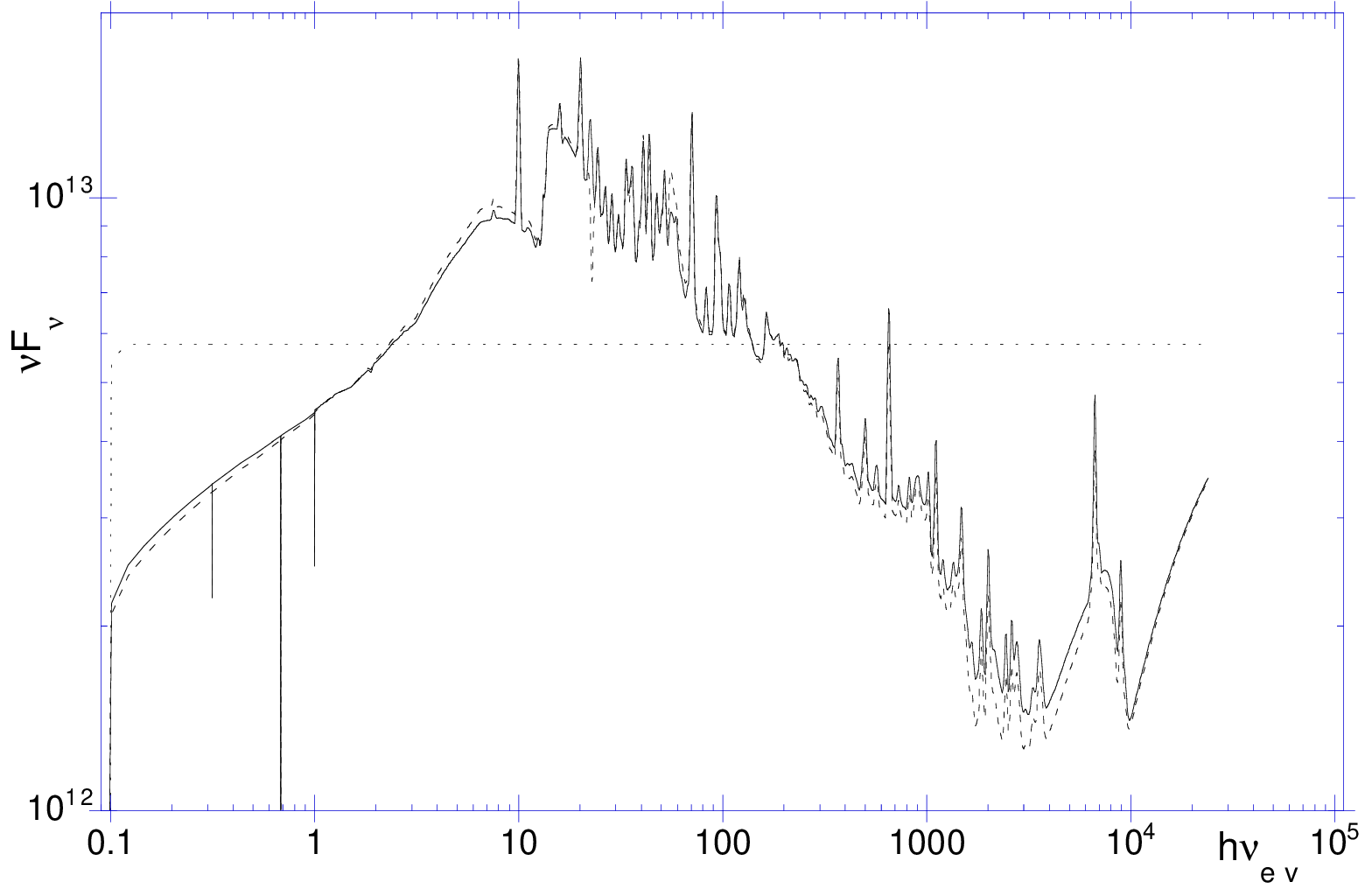}
\caption
{Reflected spectrum, with (full line), and without (dashed lines) 
ALI transfer for the continuum, for the reference model defined in Section 
4; dotted line: incident spectrum; the spectrum is displayed with a spectral resolution of 30.}
\label{GssavecALIspeAbis}
\end{figure}

We are concerned by media with a continuum optical thickness larger than 
unity in a range of frequencies rich in emission or in absorption
lines, and having a very inhomogeneous structure. Moreover in the hot 
photoionized 
layers, the absorption coefficient is weak at all wavelengths and 
Thomson diffusion is dominant.
On the contrary  absorption dominates and must be carefully treated to 
compute the spectrum emitted by the non illuminated cold part of the 
medium.
Radiation transfer requires therefore some attention.

The transfer equation is

\begin{equation}
\mu {dI \over dz} = -(\kappa + \sigma )I + \sigma J +\eta, 
\end{equation}
 where $z$ is the distance to the illuminated surface, 
$\mu$ is the cosine of the angle between the normal and the light ray,
$\kappa$ is the absorption coefficient (for the continuum it is due 
to photoionizations and free-free 
transitions), 
$\sigma$ is the diffusion coefficient (here Thomson scattering),
$\eta$ is the emissivity (due to radiative recombinations and free-free 
emission), 
and $J$ is the mean intensity, equal to $ \int I_{\nu} d\mu/2 $ 
(all these quantities are local, $I$ depends on the direction and on
the frequency $\nu$, and J, $\kappa$ and $\eta$ depend on the frequency). 

Introducing the source function $S$ and the optical depth $\tau$, 
Eq. 1 can be written:

\begin{equation}
\mu {dI \over d{\tau}} = -I + S 
\end{equation}
 with

\begin{equation}
 S = {{\sigma J +\eta} \over {\sigma +\kappa}}, 
\label{eq-trans3} \end{equation}
which can be separated between J-dependent 
and J-independent components:

\begin{equation}
 S = (1 - \epsilon) J +\epsilon B.
\end{equation}

The formal solution is:

\begin{equation}
J(\tau) = {1 \over 2} \int d\mu  \int S(t) e^{-(t- \tau)/ \mu } dt/ \mu.
\end{equation}

\subsection{Transfer of the continuum: the ALI method}

\begin{figure}
\plotone{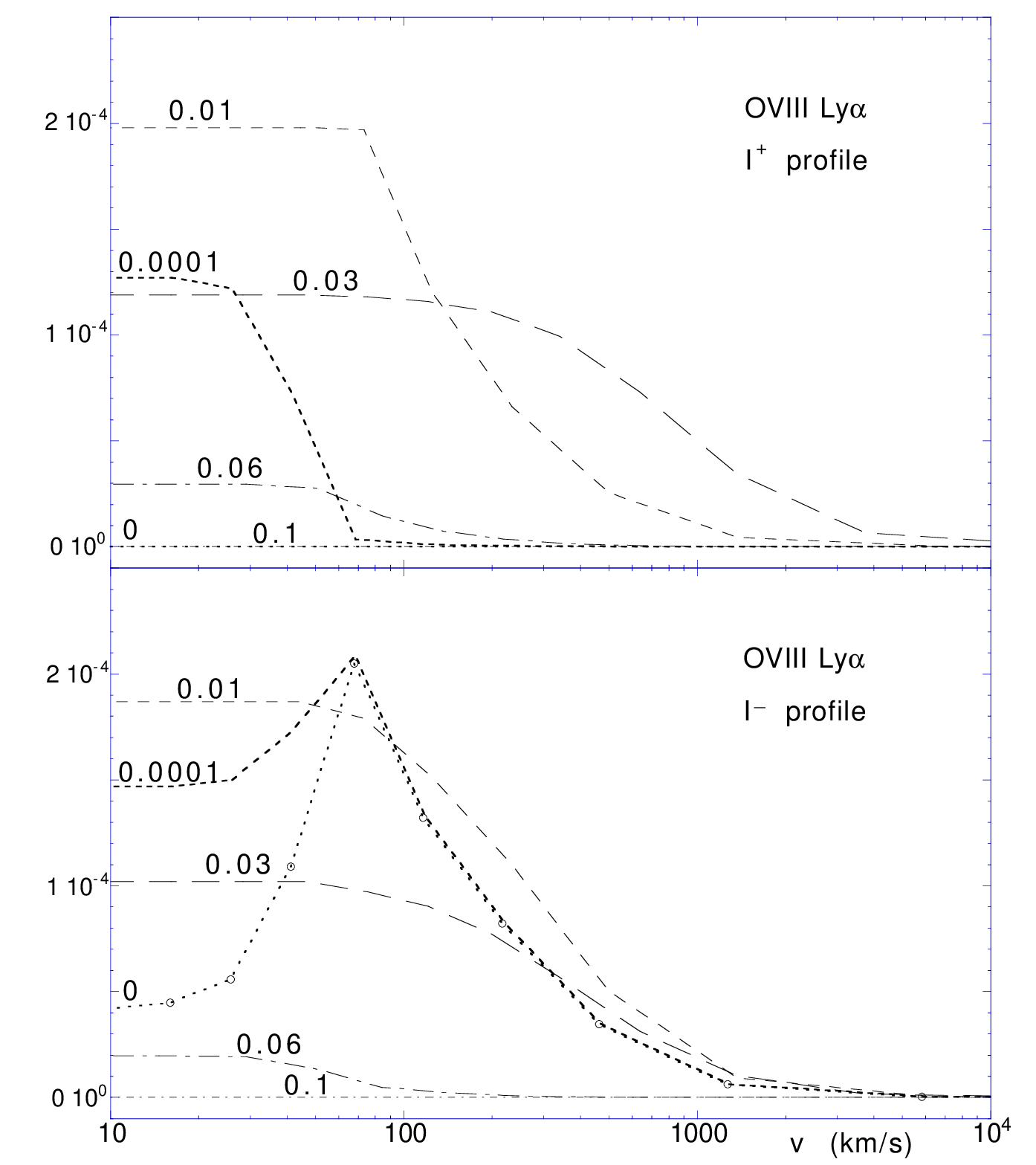}
\caption
{Profiles of $I_{\nu}^{-}$ and $I_{\nu}^{+}$ for OVIII Ly${\alpha}$, 
for the reference model defined in Section 4. The curves are labelled by 
the value of $z/H$.}
\label{fig-GprofO8Lya}
\end{figure}

In the previous version of the code (Dumont et al. 2000) we used the same 
method for 
line and continuum transfer, described in Section 3.2. Presently 
the transfer of the continuum is performed with the
{\bf Accelerated Lambda Iteration} method (ALI), which was set up with the 
help of
subroutines kindly provided to us by F. Paletou (cf.  Auer \& Paletou 1994, 
and Kunasz \& Auer 1988).

In this method $J(\tau)$ is 
written with the help of an operator $\bf\large {\Lambda}$:

\begin{equation}
J(\tau) ={ \bf\large {\Lambda} }[ S(t) ].
\end{equation}
ALI uses an operator splitting, described first by Cannon (1973),
then by Olson et al. (1986):

\begin{equation}
{\bf\large{\Lambda}} = {\bf\large{\Lambda^*}} + ({\bf\large{\Lambda}} - 
{\bf\large{\Lambda^*}})  {\rm \ \  and \ \ }  S = S^* + \delta S
\end{equation}
where $\bf\large{\Lambda^*}$ is an approximate $\bf\large{\Lambda}$ operator 
chosen to facilitate the inversion, and $S^*$ has been calculated 
in the previous iteration. One obtains: 

\begin{equation}
\delta S = [1 - (1-\epsilon) {\bf\large{\Lambda ^*}}]^{-1} [S - S^*].
\end{equation}

We use here a parabolic interpolation in the evaluation of the 
integral giving the intensity from the source function, and an 
acceleration type Ng (in this method, at every fourth iteration 
a least squares extrapolation is made in order to reduce the residuals,
cf. Ng 1974).

To perform this computation, one needs to know the opacities and
emissivities at each frequency as functions of $z$, hence the temperature 
and the populations at each depth, so an iteration procedure is required.

\begin{itemize}

\item For each layer, starting from the illuminated side, the ionization 
and thermal balance equations are solved by iteration; 
the temperature, the opacity and the emissivity are computed;

\item when the back side of the cloud is reached, the transfer is solved 
with the ALI method, and $J_{\nu} (z)$ is computed;

\item the whole calculation is repeated until convergence with the new 
values of $J_{\nu} (z)$. It is stopped when the energy balance is achieved 
for the whole slab (i.e. when the flux entering on both sides of 
the slab is equal to the flux coming out from both sides).
This method is almost as slow as the previous one, albeit more secure.

\end{itemize}

Fig. 2 shows that the difference in the spectrum  computed with the previous 
transfer method and with ALI is very small. It is certainly due to the fact that the 
continuum optical thickness is never very large (at most 
10$^4$). This is not the case at the center of intense 
lines, which are typically 100 times thicker. We are presently implementing 
ALI for the line transfer and we 
expect to find more important differences for a few lines like OVIII 
L$\alpha$ which are not completely converged even after 1000 iterations 
(cf. Dumont et al. 2000).

\subsection{Line transfer}

Since ALI is presently not implemented for the line transfer, it is treated
 like in the previous version with a simple method
based on the Eddington 
two-stream approximation. In this approximation the transfer equations are written:

\begin{eqnarray} {1 \over \sqrt{3}}{dI_{\nu}^{+} \over dz}
&=& -(\kappa_{\nu} +{\sigma  \over 2})I_{\nu}^{+}+{\sigma  \over
2}I_{\nu}^{-}+\epsilon_{\nu} \\
\nonumber {-1 \over \sqrt{3}}{dI_{\nu}^{-}
\over dz} &=& -(\kappa_{\nu} +{\sigma  \over
2})I_{\nu}^{-}+{\sigma  \over 2}I_{\nu}^{+}+\epsilon_{\nu}.
\label{eq-trans1}
\end{eqnarray}

The mean intensity  $J_{\nu}$ is equal to $(I_{\nu}^{+}+I_{\nu}^{-}) / 
2$, 
and the flux $F_{\nu}$ defined as usual by
$ \int I_{\nu} \cos{\theta}\ d\omega $ is equal to
$(I_{\nu}^{+} - I_{\nu}^{-})\ 2\pi /\sqrt{3}$.
The ``reflected" flux is equal to $I_{\nu}^{-} (0)\ 2\pi /\sqrt{3} $,
and the outward flux to $I_{\nu}^{+} (H)\ 2\pi /\sqrt{3}$.
The optical depth and the total optical depth are defined  as:

\begin{equation} 
\tau _{\nu}(z)=\int_{0}^{z} \sqrt{3}(\kappa_{\nu} +\sigma )dz'\ \ \
{\rm and}\ \  T_{\nu}=\tau_{\nu}(H)
\label{eq-trans2}
\end{equation}

\noindent where $H$ is the total thickness of the slab.
For $I_{\nu}^{+}(z)$ the formal solution of the transfer equations
between $z-\delta z$ and $z$ is:

\begin{equation}
I_{\nu}^{+}(z) = I_{\nu}^{+}(z-\delta z) e^{-\delta\tau_{\nu}}
 + e^{-\tau_{\nu}}\int_{\tau_{\nu}-\delta\tau_{\nu}}^{\tau_{\nu}}
S_{\nu}(t) e^{+t} dt.
 \label{eq-trans5} \end{equation}

We approximate the source function $S_{\nu}(t)$ 
by a constant in the interval $z, z+dz$. It leads to:

\begin{equation}
I_{\nu}^{+}(z) = I_{\nu}^{+}(z-\delta z) e^{-\delta\tau_{\nu}} +
 {{1-e^{-\delta\tau_{\nu}}}\over 2}[S_{\nu}(z-\delta z) + S_{\nu}(z)]
 \label{eq-trans6} \end{equation}
and:

\begin{equation}
I_{\nu}^{-}(z) = I_{\nu}^{-}(z+\delta z) e^{-\delta \tau_{\nu}}+
{1-e^{-\delta\tau_{\nu}}\over 2}[S_{\nu}(z+\delta z)+S_{\nu}(z)].
 \label{eq-trans7} \end{equation}

We calculate  $S_{\nu}(z)$ and the mean intensity $J_{\nu}(z)$ from
Eq. 3 and from:

\begin{eqnarray}
J_{\nu}(z) &=& {I_{\nu}^+(z-\delta z) + I_{\nu}^-(z+\delta z)
\over 2} e^{-\delta \tau_{\nu}} \\
\nonumber
&+& {(1-e^{-\delta\tau_{\nu}})\over 4}
[S_{\nu}(z-\delta z) + 2S_{\nu}(z) + S_{\nu}(z+\delta z)],
\label{eq-trans10} 
\end{eqnarray}
assuming a constant $\delta z$ (actually it is not constant so the
formulae are more complicated but do not deserve to be given here).

\begin{figure}
\plotone{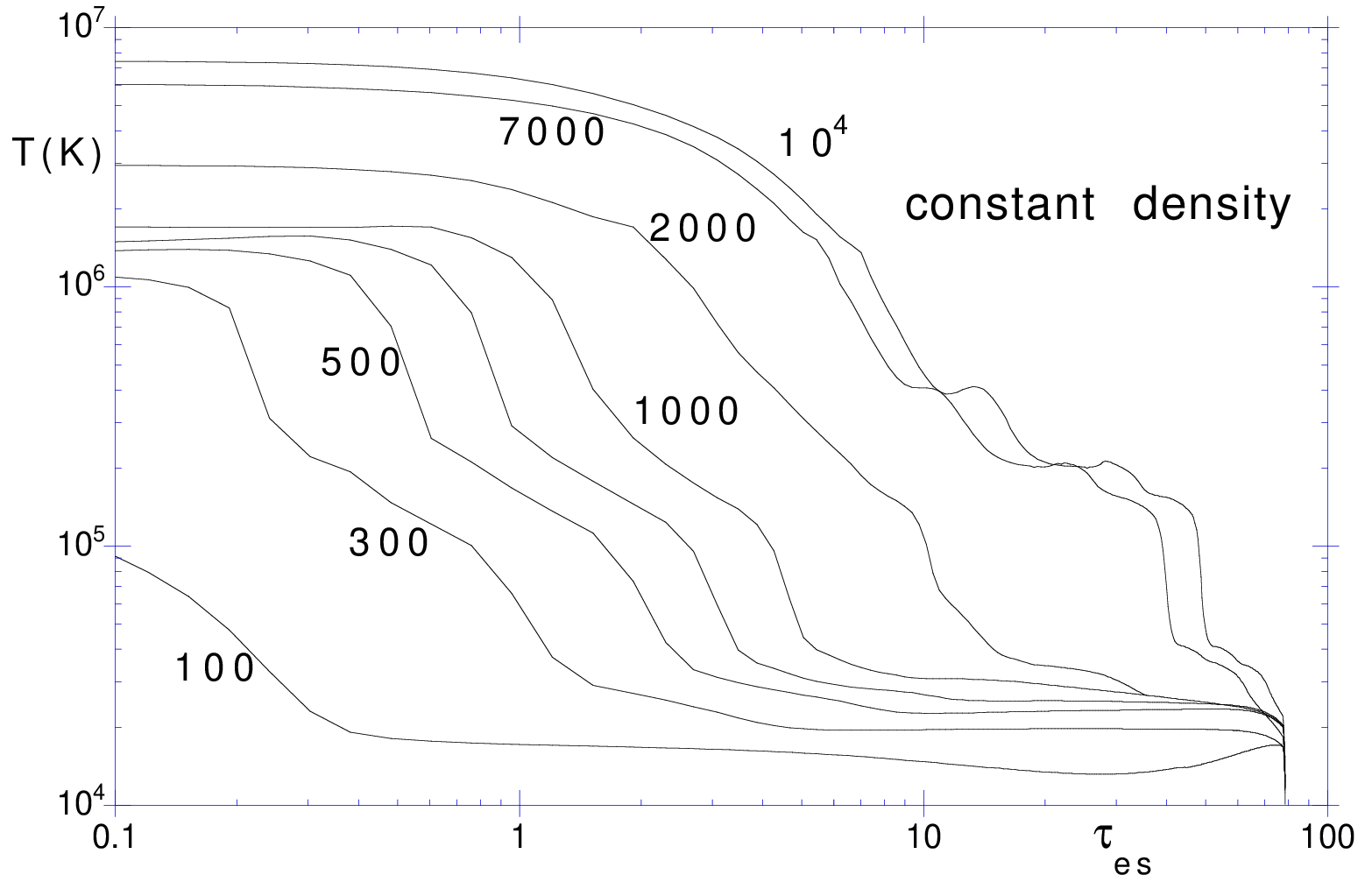}
\caption
{Temperature versus depth for different values of $\xi$ in erg
cm s$^{-1}$, the other parameters being the same as in the reference 
model.}
\label{fig-Gn12c26-csi-tem}
\end{figure}

The procedure is then the following:

\begin{itemize}

\item for each layer, after computation of the temperature and 
the source function, $I_{\nu}^+(z)$ is transferred through the layer
according to Eq. 12, while $S_{\nu}(z)$ and $\tau_{\nu}(z)$ are
buffered for each layer and each frequency;

\item  when the back side of the cloud is reached, new values of
$I_{\nu}^{-}(z)$  are calculated, starting from the back side
where $I_{\nu}^{-}(H)$ is given by Eq. 12, and using Eq. 13 with the previous values
of $S_{\nu}(z)$ and $\tau_{\nu}(z)$ (actually we use the values given by
several previous iterations to accelerate the procedure with a Ng method).

\end{itemize}

This method requires a large number 
of iterations to get a complete convergence of the model, and 
one cannot be sure to have reached the correct value. It is why we are
now changing for the ALI method.

We assume presently complete 
redistribution in the lines, and partial redistribution 
in intense lines is mimicked by assuming complete redistribution 
in a pure Doppler profile. It will be possible to treat partial 
redistribution 
with the ALI method.

A line profile is represented by a Voigt function at several 
frequencies distributed 
symmetrically around the line center. The 
frequencies in the profile are chosen to take into 
account the fact that the Doppler width may decrease by one order of 
magnitude inside the slab, 
owing to large temperature variations.
Moreover, to solve the statistical equilibrium equations one needs to know 
the total line intensity weighted by the profile 
$ \int J_{\nu} \phi(\nu) d\nu $, while to compute the thermal and 
ionization equilibria one simply needs the line intensity
integrated over frequencies  $ \int J_{\nu} d\nu $.
For intense lines, the first integral is dominated by the Doppler 
core, and the second by the wings. The set of frequencies chosen 
to describe the line must therefore cover at least two orders of 
magnitude.  The integrals 
over the line profiles are achieved with a 15-point 
Gauss-Legendre quadrature. We also assume that the lines do not 
overlap, which is valid as far as multiplets are treated globally, 
and because the number of lines taken into account in TITAN is relatively 
small (presently 580, including subordinate lines in the optical-UV range).  


\begin{figure}
\caption
{Reflected and outward spectra for different values of $\xi$ in erg cm s$^{-1}$,
the other parameters being the same as in the reference model; 
the spectra are displayed with a spectral resolution of 100.}
\label{fig-Grefl-sort-csi}
\end{figure}


\begin{figure}
\caption
{Reflected and outward spectra for different values of the column density 
in cm$^{-2}$, 
for $\xi=100$ erg cm s$^{-1}$, the other parameters being the same as 
in the reference model; the spectra are displayed with a resolution of 100.
 Dotted line: the incident 
spectrum.}
\label{fig-Gx2R100-ref-sort-col}
\end{figure}

As shown by Fig. 3, the profiles of  $I_{\nu}^{-}$ and  
$I_{\nu}^{+}$ for OVIII Ly${\alpha}$ vary
considerably from the surface to the deepest layers.

\section{Some results of TITAN}

	We shall call ``reference model" a model with the 
	following characteristics:

\begin{itemize}

\item  it is a  plan-parallel slab with a constant density, equal to 10$^{12}$ 
 cm$^{-3}$,

\item  its column density equals 10$^{26}$ cm$^{-2}$ (corresponding 
to a Thomson thickness $\tau_{\rm es}$ of 80), 

\item it is illuminated on one side by an incident power law continuum
proportional to $\nu^{-1}$ from 0.1 eV to 100 keV,

\item  the ionization parameter is $\xi=10^3$ erg cm s$^{-1}$,

\item  there is no illumination on the  other side,

\item the abundances are  (in number): H: 1, He: 0.085, 
C: 3.3 10$^{-4}$,  N: 9.1 10$^{-5}$, O: 6.6 10$^{-4}$, 
Ne: 8.3 10$^{-5}$, Mg: 2.6 10$^{-5}$, Si: 3.3 10$^{-5}$, S:
1.6 10$^{-5}$, Fe: 3.2 10$^{-5}$.

\end{itemize}

	In several examples given below, we also use a smaller value 
of the ionization parameter, $\xi=10^2$ (corresponding to a value of the 
usual ionization parameter $U$ 
equal to 0.88), allowing an easier comparison 
with computations made with CLOUDY or with XSTAR. 


Figs. 4 and 5 show the results corresponding to different values of 
the ionization
 parameter, the other parameters being the same as in the reference model. 
These results were obtained with the previous version of the code. Indeed
 each model takes a long time to run, and all models have still
 not been recalculated with the 
ALI method. However, we have checked in several cases that the results 
obtained with the new version differ 
very slightly from the previous ones.

Fig. 4 displays the temperature versus the depth. Close to the surface the  
temperature is high and reaches the Compton temperature for the highest value
 of 
$\xi$. The extension of this region (called 
the ``hot skin") increases with the value of $\xi$.  Its optical 
thickness  reaches 
$\tau_{\rm es} \sim 10$ for $\xi=10^4$. As we shall see later, this is quite 
different from the case of constant pressure, where the thickness of this 
zone is always limited to $\tau_{\rm es} \le 1$. This hot skin is almost 
absent for low values of $\xi$. Below the hot skin the temperature 
decreases smoothly, again at
contradistinction with the case of a constant pressure. This decrease is 
followed by a region of a quasi constant 
temperature, which is in a state of ``quasi-LTE", i.e. each 
process important for the energy equilibrium (lines, free-bound from and 
onto ground levels, free-free) is exactly balanced, due to the high 
optical thickness and relatively high density. However the
region is not in LTE, in the sense that the fractional ionic abundances do 
not satisfy Saha equations. We shall discuss this issue in more detail in 
the next section. 

On Fig. 5
 are displayed the ``reflected'' spectrum emitted by the illuminated 
 side of the slab, and the ``outward'' spectrum emitted by the back side. 
Note that the outward spectrum is distinct from the 
transmitted one, as it includes also {\it emission} (actually our 
transfer method does not allow to separate 
emission and transmission, contrary to approximate methods using for instance ``on 
the spot" approximation). The outward spectrum lies only in the optical 
near UV range, owing to the low temperature of the back side (cf. Fig. 
4). This is due to the large column density of the model which does not 
allow any transmission in the X-ray range. 
Note the strong edges in emission (including a Balmer edge), and the cut 
off of the spectrum at a few tens of eV, due to absorption by HeII 
and by low ionization heavy element species. The reflected spectrum 
shows a characteristic shape with a smooth ``powerlaw" soft X-ray excess, 
and displays
 many lines and ionization edges whose intensity decreases with increasing 
values of $\xi$, owing to the increasing influence of pure reflection when 
the   
ionization state increases. Note that the Lyman discontinuity is in 
emission for low $\xi$ and in absorption for high $\xi$.

Fig. 6  illustrates the influence of the column density on the reflected and 
on the outward spectra for $\xi=100$, the other parameters being the same as 
in the reference model. First we
 see that the reflected 
spectrum ``saturates" above 10$^{25}$ cm$^{-2}$, because all the available 
photons have been absorbed and no longer succeed in escaping from the 
illuminated side. Second, the optical-UV part of the outward spectrum  
saturates above 10$^{25}$ cm$^{-2}$, but not the X-ray part, which is due 
to transmission. As the column density decreases, the slab becomes 
more transparent and less emissive, so the outward spectrum becomes 
basically only the transmitted spectrum, while the reflected 
spectrum decreases in 
intensity. Note that 
the ``saturation" of the reflected spectrum would occur at a larger value of
 the column density for a
larger $\xi$, as the hot reflecting skin becomes thicker. It is thus 
{\it necessary to take into account deeper layers, exceeding a Thomson 
thickness $\tau_{\rm es}$ of few units, 
to compute correctly the reflected spectrum in the UV range
 for $\xi \ge 10^3$, when 
dealing with constant density slabs}. This has always been overlooked in 
the past.

\subsection{Influence of the excited levels}

In the first version of the code, non hydrogen-like ions were treated with a 
rough approximation: the populations of the excited levels were 
computed separately using a two-level 
approximation, i.e. interlocking between excited levels was neglected, as 
well as ionizations from these levels. Moreover only a very small number of 
levels (at most four) were taken into account. However recombinations 
onto excited states were taken into account in the ionization equilibrium, and
 the 
transfer of these photons was treated in an approximative way, according 
to an escape probability formula proposed by Canfield \& Ricchiazzi
(1980). Each recombination was also assumed to produce a 
resonant photon after cascades. 

Hydrogen and hydrogen-like ions were - and are - treated as 6-level atoms.
 Levels 2s and 2p 
are treated separately, while full l-mixing is assumed for higher levels. All 
processes including collisional and radiative ionizations and recombinations 
are 
taken into account for each level (cf. Mihalas 1978). 
 In the previous version, recombinations onto 
levels higher than five were not taken into account, which amounted 
assuming that the higher levels were in LTE with the continuum. It was true
for 
hydrogen in the conditions 
in which this code is presently used, but not for other hydrogen-like ions. 
We have therefore recently added recombinations
 onto higher levels, and we intend to implement 
``fictive" levels to take ionizations from upper levels into account as it is done 
for instance in CLOUDY or XSTAR. 
Level populations are obtained as usual by matrix inversion.

The rough treatment of non  hydrogen-like ions could 
have several consequences. 

\begin{figure}
\plotone{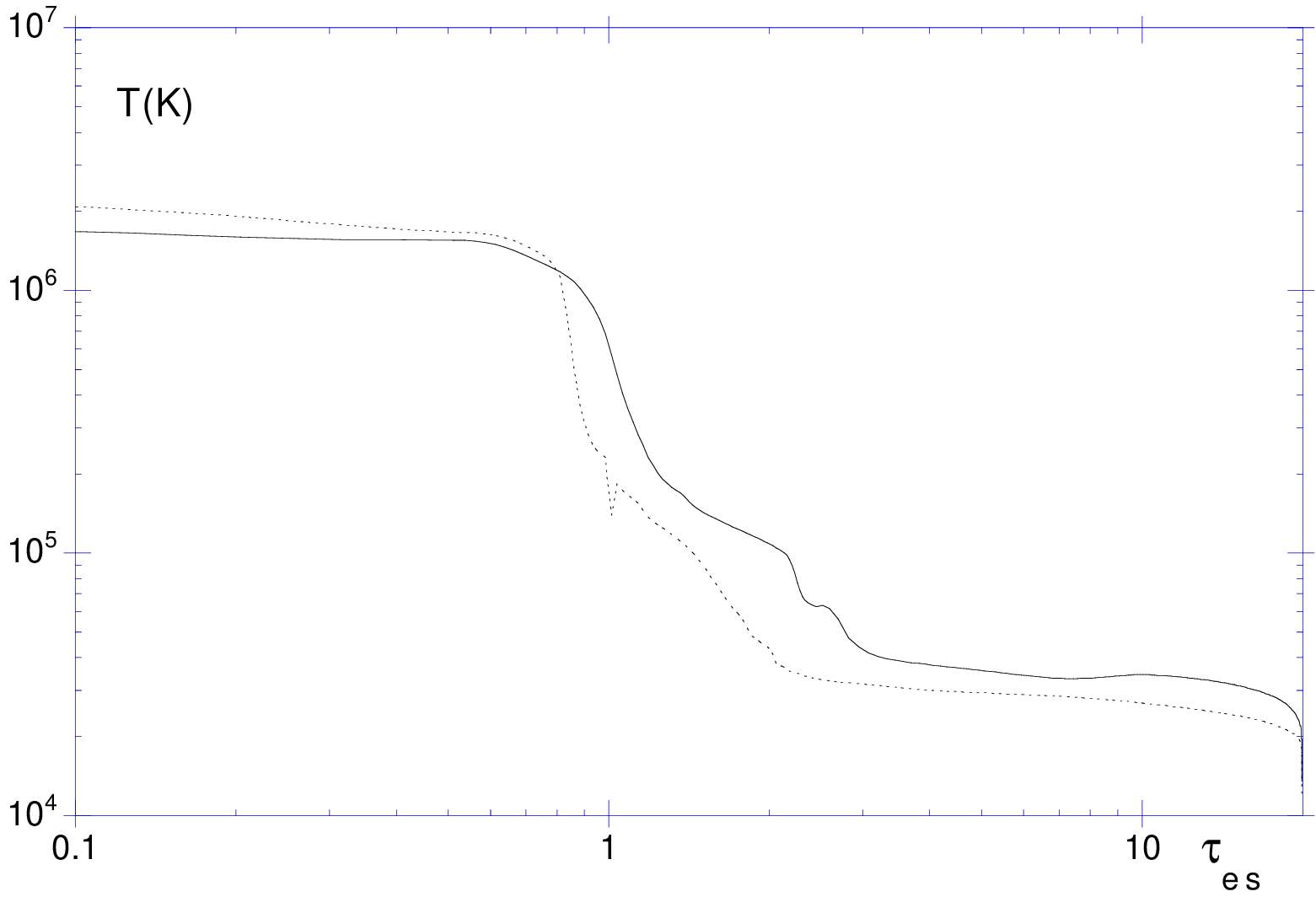}
\caption
{Comparison of the temperature 
computed with (full line) and without (dashed lines) full eight-level 
 He-like and full five-level Li-like ions,  for the reference model,
 versus the distance to the illuminated side.}
\end{figure}

\begin{itemize}
\item First 
it should {\it not predict correctly the details of the 
line spectrum}, since it neglects subordinate lines and even important 
resonance lines. 
\item Also it can  lead to a {\it bad estimation 
of the energy losses} and consequently to a change of the temperature. The
 recombination 
continuum corresponds generally to a smaller absorption coefficient than the neglected 
lines, and therefore can escape more easily from the deep layers. It would correspond 
to an overestimation of the losses and an underestimation of the 
temperature. But  other parameters, such as the 
ionization state, play also a crucial role in the energy balance, so the net effect 
is difficult to predict. 
Fortunately in these models cooling due to lines is not 
predominant except in some regions. 
\item It can lead to important errors in the {\it fractional ion 
abundances}. 
This effect is ``self-cumulating", as an increased proportion of a 
given ion leads to a stronger ionization edge, which in turns inhibits 
further photoionizations. So again the net effect is  not predictable.
\item Finally recombinations onto excited levels are 
not balanced by ionizations from these levels. It is not important in 
conditions prevailing in a photoionized slab far from equilibrium (like for 
instance in the hot skin of an irradiated accretion disk), as the 
lowest atom levels are strongly overpopulated with respect to LTE. But this 
treatment is not 
valid in the innermost layers of a disk atmosphere, where the density is 
very high so complete LTE is almost 
reached.

\end{itemize}
 
We have recently introduced  
interlocked multi-level computations for He-like and Li-like ions, in order
to obtain a better description of the spectrum, and 
to treat cases closer to LTE. 
However we are for the moment restricted to 8 levels for He-like ions and 5 
for Li-like ones, which is not sufficient close to LTE.  We take into 
account 
recombinations onto the levels higher than 8 for He-like ions 
(respectively 5 for Li-like), adding them as a contribution to the 
recombination of the highest level. This treatment is 
similar to the old one, except that it takes into account many 
subordinate lines and ionizations from excited levels, and it is valid as far 
as LTE is not reached for the highest levels and ionizations are dominated 
by the ground level. Like for hydrogen-like ions 
we plane to add several 
other levels, and to mimic the contributions of the higher levels.

We mention also that radiative recombinations are computed differently according 
to the relative 
values of $kT$ to the photon energy bin, in order to get an accurate 
frequency dependence.

Figs. 7 to 10 illustrate the influence of the new treatment with respect to the 
old one, for the reference model (except that the thickness is only 
$\tau_{\rm es}= 25$ instead of 80; we have already mentioned that
above this value nothing varies anymore for $\xi=10^3$, cf. Fig. 4). In this model the 
level populations and the ionization degrees are far from LTE, even in the 
deepest layers. It is quite comparable to a classical photoionized 
model, where excited levels are populated by radiative recombinations and 
depopulated by deexcitations onto lower levels. Note that these results 
are obtained with the new transfer treatment.

Fig. 7 shows the new temperature profile. Except at the surface $T$ is increased with 
respect to the previous computation. The change is quite important, 
reaching a factor two in the 
transition layer. Fig. 8 shows the change of fractional abundances of Hydrogen, Helium, and 
Iron. It is relatively important in the hot skin for HeI, but the 
abundance of this ion is completely negligible in this region, so it should 
not induce any change in the emitted spectrum around the ionization edge, 
i.e. 20 eV. The result of a 
computation with full He-like and Li-like atoms, but without additional 
recombinations on the highest level, is also shown. Except precisely for 
HeI in the hot skin, the results with and without the additional term are 
quite similar. 

\begin{figure}
\plotone{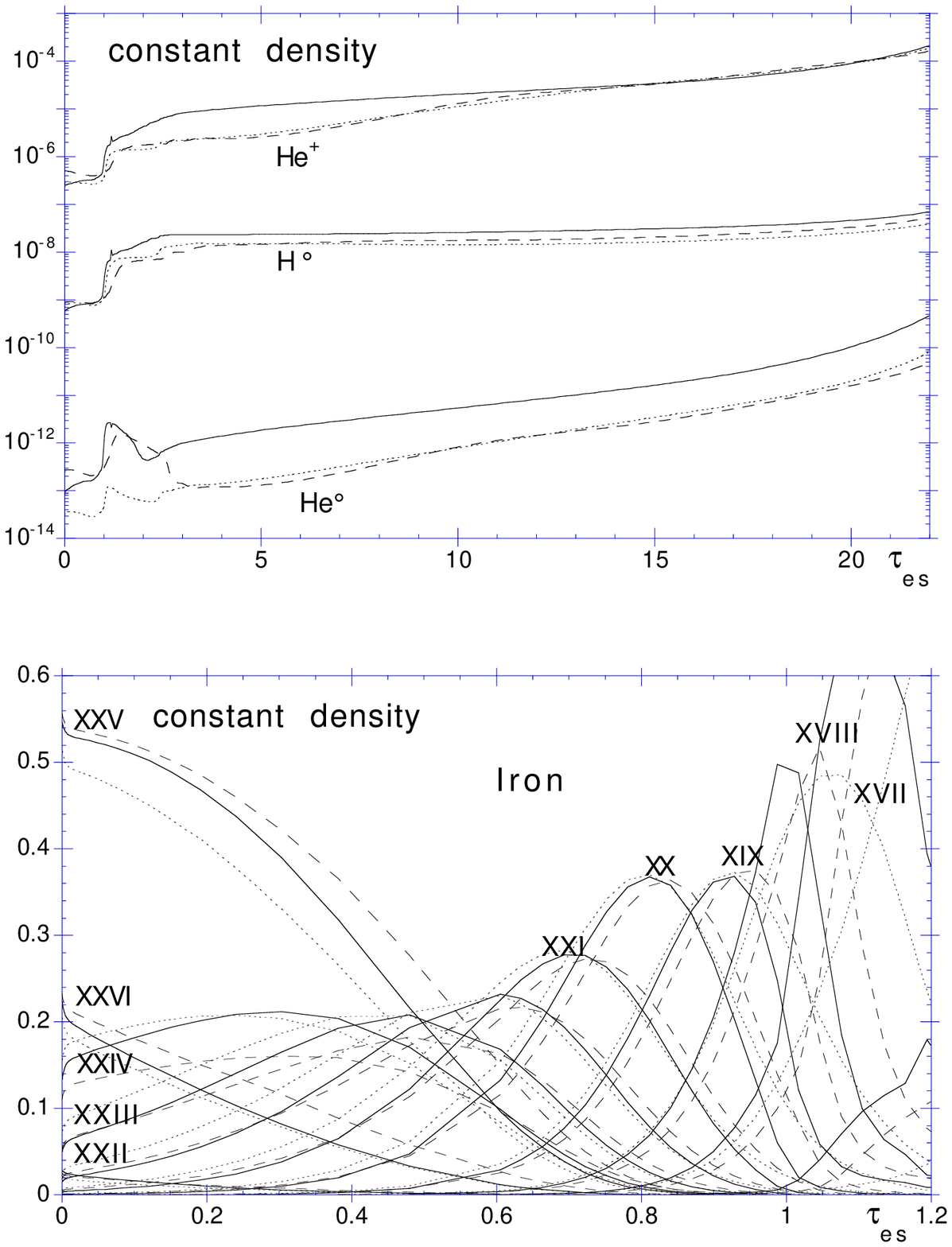}
\caption
{Fractional abundances of Hydrogen, Helium and Iron,
computed with (dotted line) and without (full lines) 
full eight-level 
 He-like and full five-level Li-like ions,  
for the reference model,
 versus the distance to the illuminated side.   The result of a 
computation with full He-like and Li-like atoms, but without additional 
recombinations on the highest level, is also shown (dashed lines).}
\end{figure}

\begin{figure}
\plotone{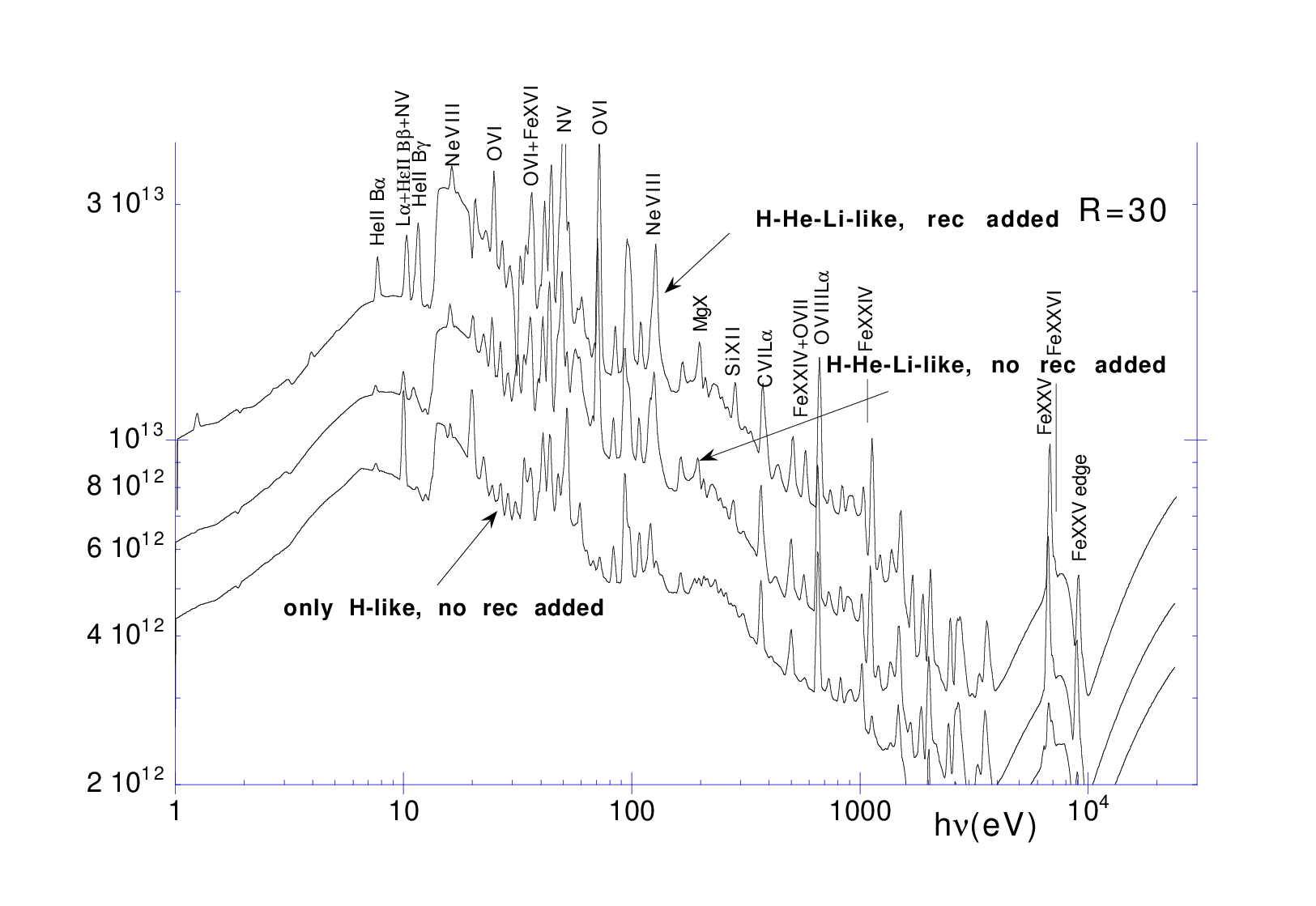}
\caption
{Reflected spectrum computed with and without 
full eight-level 
He-like and full five-level Li-like ions,  for the reference 
model.  The result of a 
computation with full He-like and Li-like atoms, but without additional 
recombinations on the highest level, is also shown. The spectra are 
vertically shifted for a better reading.
They are displayed with a resolution of 30. A few intense lines are 
identified.}
\end{figure}

\begin{figure}
\plotone{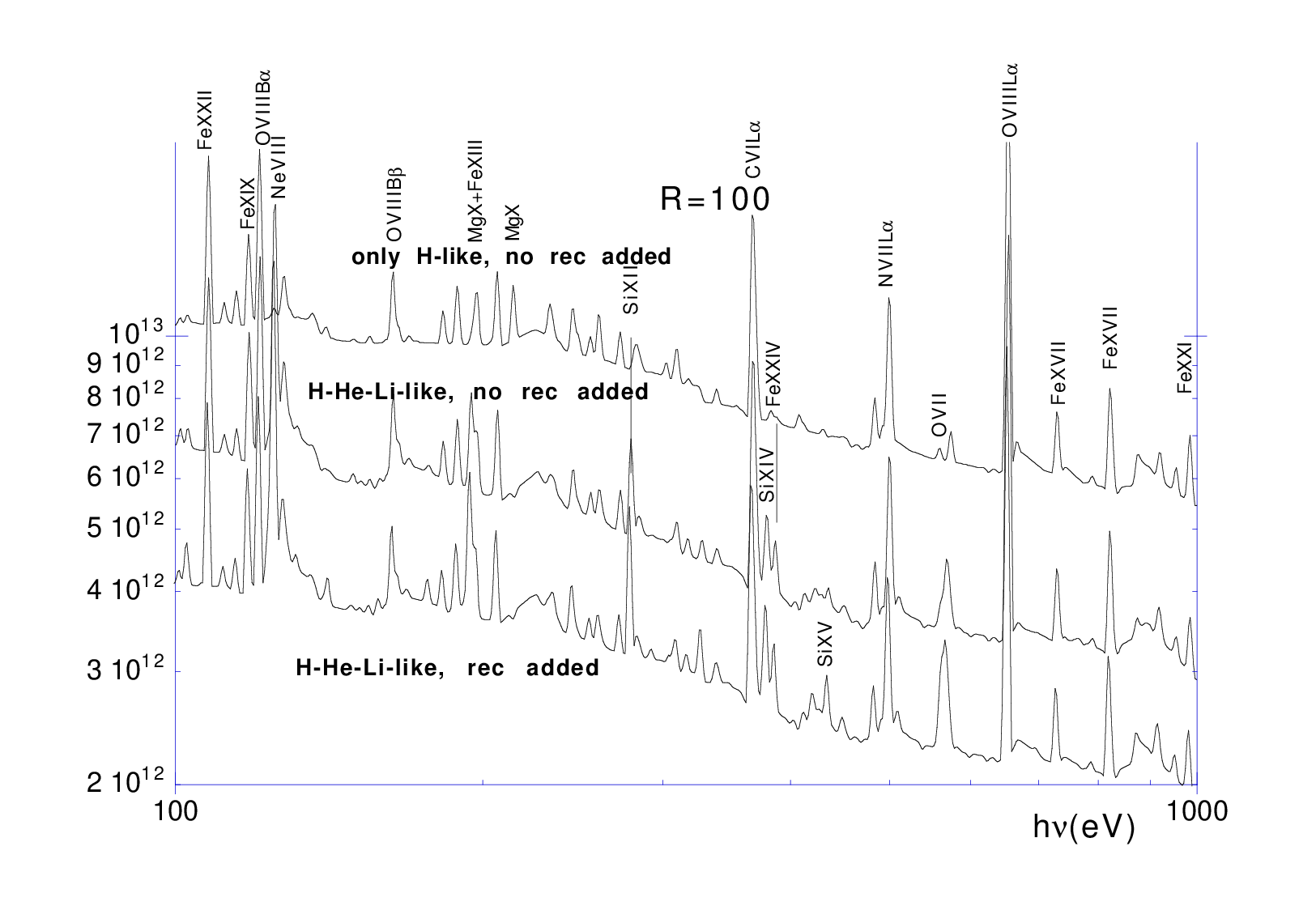}
\caption
{Reflected spectrum computed with and without 
full eight-level 
He-like and full five-level Li-like ions,  for the reference 
model.  The result of a 
computation with full He-like and Li-like atoms, but without additional 
recombinations on the highest level, is also shown. The spectra are 
vertically shifted for a better reading.
They are displayed with a resolution of 100. A few intense lines are 
identified.}
\end{figure}

Figs. 9 and 10 display the reflected spectrum for this model, computed with the old 
and the new atomic treatment, with a low resolution in the 
whole energy range (Fig. 9), and with a higher
 resolution in a smallest but interesting energy 
range (Fig. 10). 

The spectral distribution does not differ much for the two treatments. This is 
not the case of the detailed features, as it can be seen on Fig. 10. 
The flux of the main lines of H-like ions in the reflected
spectrum are almost not changed (which is expected since the treatment 
is the same). 
For the He-like and Li-like ions, 
the total flux for all lines is multiplied by a factor 3 or 4. To 
illustrate this effect, Fig. 11 displays the fluxes of the OVI, OVII, and 
OVIII lines. Even the resonant lines differ in both treatments, as the 
energy is distributed among a much larger number of lines: for instance 
the OVIII line at 665 eV decreases by one order of magnitude with the new 
treatment. The net increase 
of line flux is simply due to the redistribution of continuum recombination 
photons into line photons through cascades, an effect which was only partly taken into 
account in the previous treatment.

Besides, we have added the result of a computation performed with the complete 
treatment of He-Li-like ions, when suppressing the recombination 
rates added on the highest level (cf. Figs. 9 and 10). This suppression does not
affect at all the spectral distribution, and only slightly the detailed features of the
 spectrum. 

The next further step is to introduce 
forbidden lines, which could be important for 
the cooling of the medium in some places.

 Note that all the 
results given in the following sections are obtained without  Li-like and He-like 
complete atom models.

\subsection{Influence of the density distribution}

\subsubsection{Constant pressure}:
\medskip

Up to now, we have presented results only for slabs of constant density. 
It is a peculiar case, as the pressure is thus varying 
according to the density and the temperature. When the pressure is imposed 
and not the density, one is faced to the well-known problem of thermal 
instability described first in the framework of AGN by Krolik,
 McKee \& Tarter (1981). If the pressure is imposed, three 
 (or even more) 
 solutions to the  Gain=Loss equation can exist, and one of these solutions 
is unstable. The
 gas has therefore to shift either to the Compton dominated hot solution at a 
 temperature close to the Compton value, or to 
 the cold solution dominated by atomic processes, at a few $10^4$ K.

\begin{figure}
\plotone{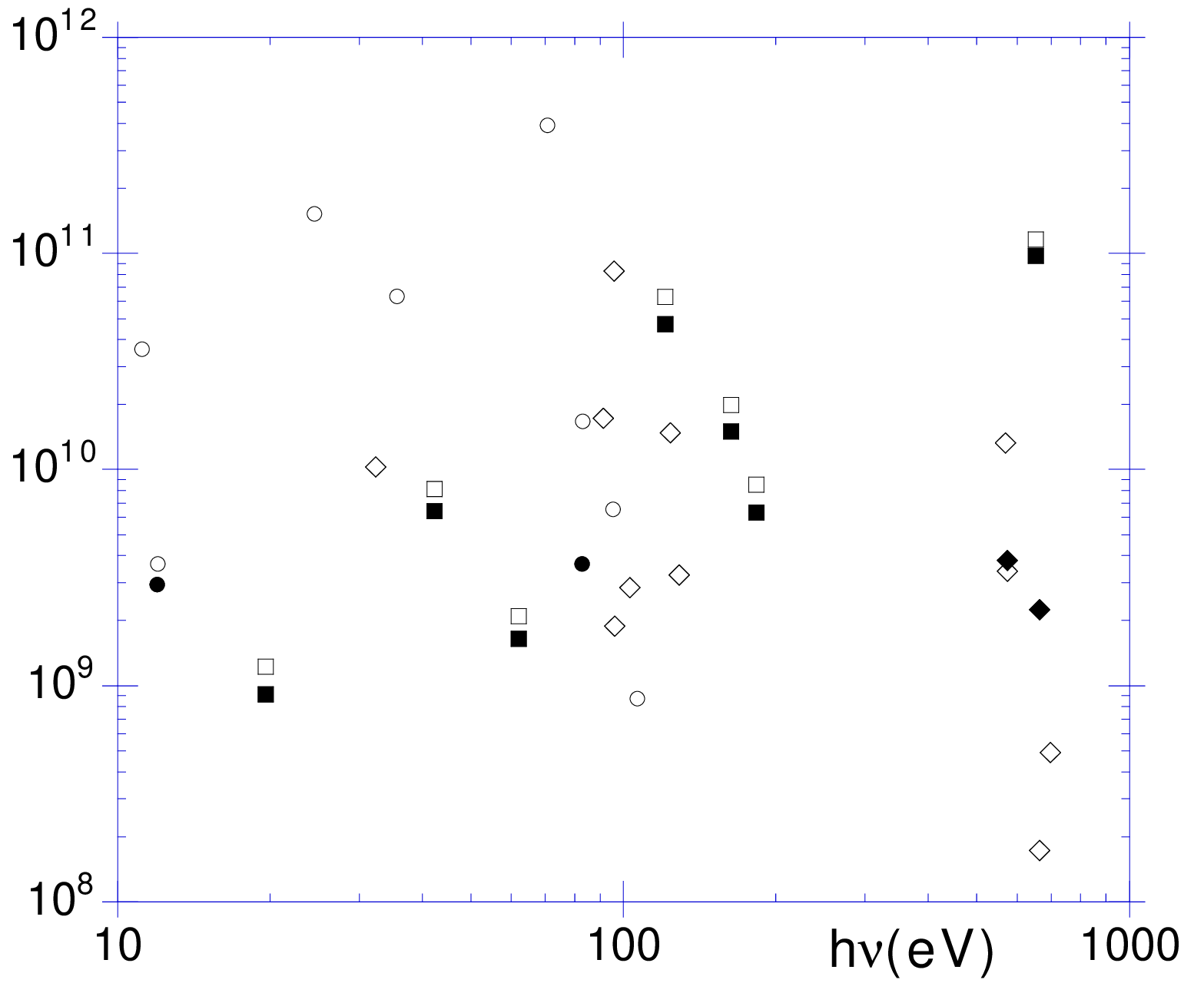}
\caption
{Line fluxes with (open symbols) and without (full symbols)
full eight-level 
 He-like and full five-level Li-like ions,  for the reference 
model.  OVI: circles; OVII: losanges; OVIII: squares.}
\end{figure}

Krolik 
 et al. were considering optically thin media with a uniform temperature, 
 whose value is given by the external irradiating spectrum. 
Here 
 we are considering a stationary case, where the temperature changes across 
 the slab
 because the mean intensity $J_{\nu}$ is modified as a function of depth. 
 Since $J_{\nu}$ is absorbed in the soft X-ray band, its
 spectral distribution becomes more dominated by UV as the depth increases, 
and such a spectrum does not lead to 
 the S-shape curve characteristic of thermal instability (cf. Krolik 
 et al. 1981). However, even if there 
 is no instability, the temperature decreases very rapidly with increasing 
 depth, as the constancy
 of the pressure leads to an increase 
 of the density, therefore to an increase of the cooling, and to a further 
 decrease of temperature
 (note that the temperature is not strictly inversely proportional to the density, 
 as radiation pressure generally exceeds gas pressure). 

\begin{figure}
\plotone{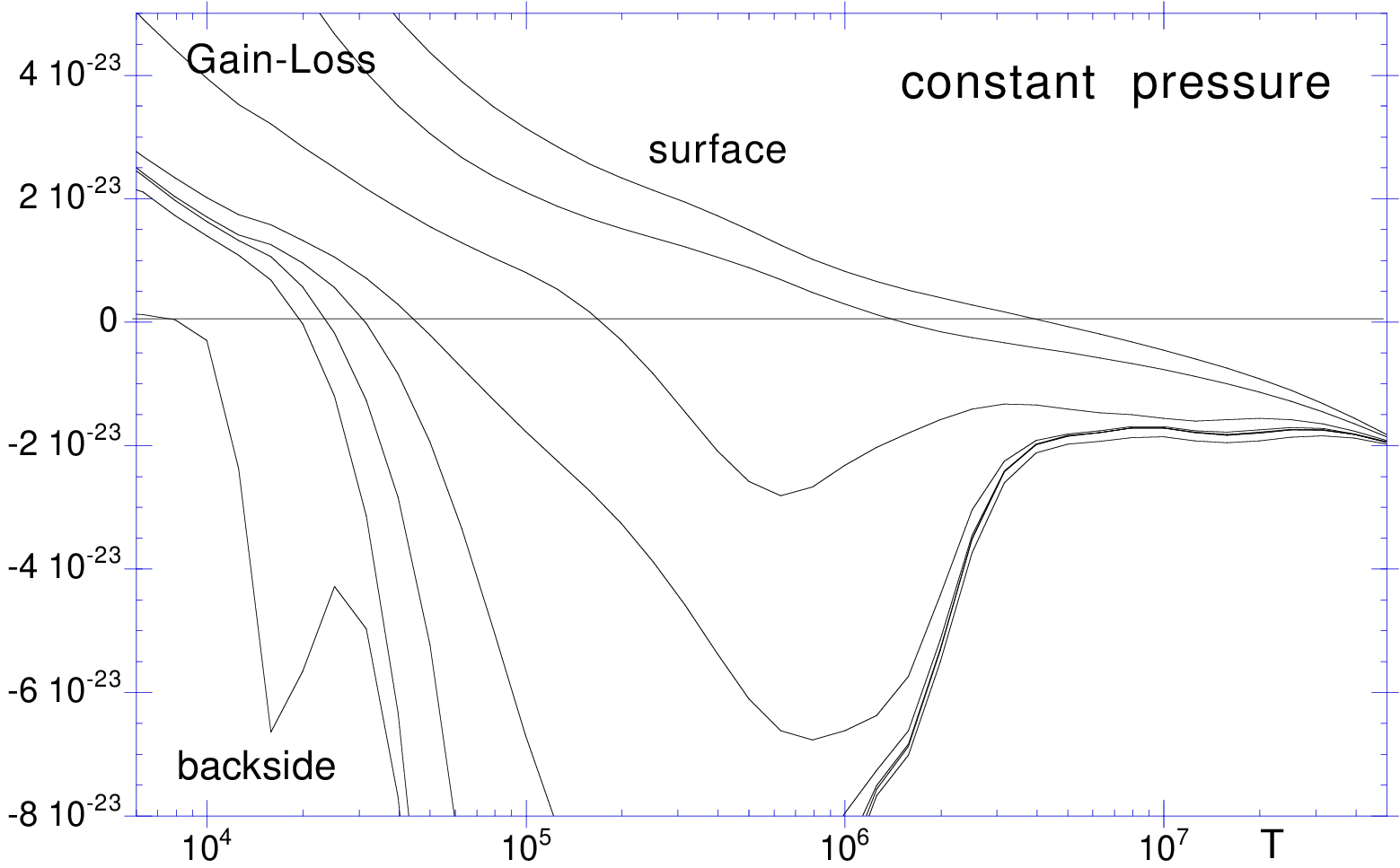}
\caption
{Gain-Loss function versus the temperature, for different layers across 
the slab, for a constant pressure model.}
\end{figure}

This is illustrated on Fig. 12 which shows the Gain-Loss function 
versus the temperature for different layers. The model is a slab with a 
constant pressure, a density at the surface equal to 10$^{11}$ cm$^{-3}$, an 
ionization parameter equal to 1000. The illuminating 
spectrum is a power law $F_{\nu} \propto \nu^{-0.7}$, chosen because such a flat spectrum 
is well known to correspond to an ``S-shape" curve with a strong instability. 
However we see that as the temperature decreases with the depth, there are never 
multiple solutions of the Gain=Loss equation. Note that it is by no means a general 
result.  In particular, a medium with a more rapid variation of the 
pressure with the depth, like given by hydrostatic equilibrium, could likely lead to a 
thermal instability, as the transition to the cold region occurs in a 
region where the spectrum is less absorbed than in a constant density 
medium (cf. Nayakshin, Kazanas \& Kallman 2000). Also if the UV band is 
absorbed before the soft X-ray one, which is the case when the ionization 
parameter is small, it should lead to a thermal instability.

Fig. 13 displays the temperature versus depth for slabs of constant 
density and slabs of constant pressure, with different ionization 
parameters. Note that it would be better to speak of ``ionizing 
flux" instead of ionizing parameter in the constant pressure case. 
Actually the comparison is performed with the same incident fluxes. 
As expected, and in contrast with the constant density case where there is a smooth
decrease of the temperature with increasing depth, the constant pressure case gives
 a sharp transition between the 
hot and dilute, and the cold and dense, parts of the slab. Moreover the 
thickness of the ``hot skin" is smaller for the same flux in the constant 
pressure case, in particular for high illumination. For instance, for 
$\xi=10^3$ the Thomson thickness of the hot skin is about unity in the 
constant density case, and only 0.4 in the constant pressure case. All 
these effects have strong consequences on the emission spectrum.


\begin{figure}
\plotone{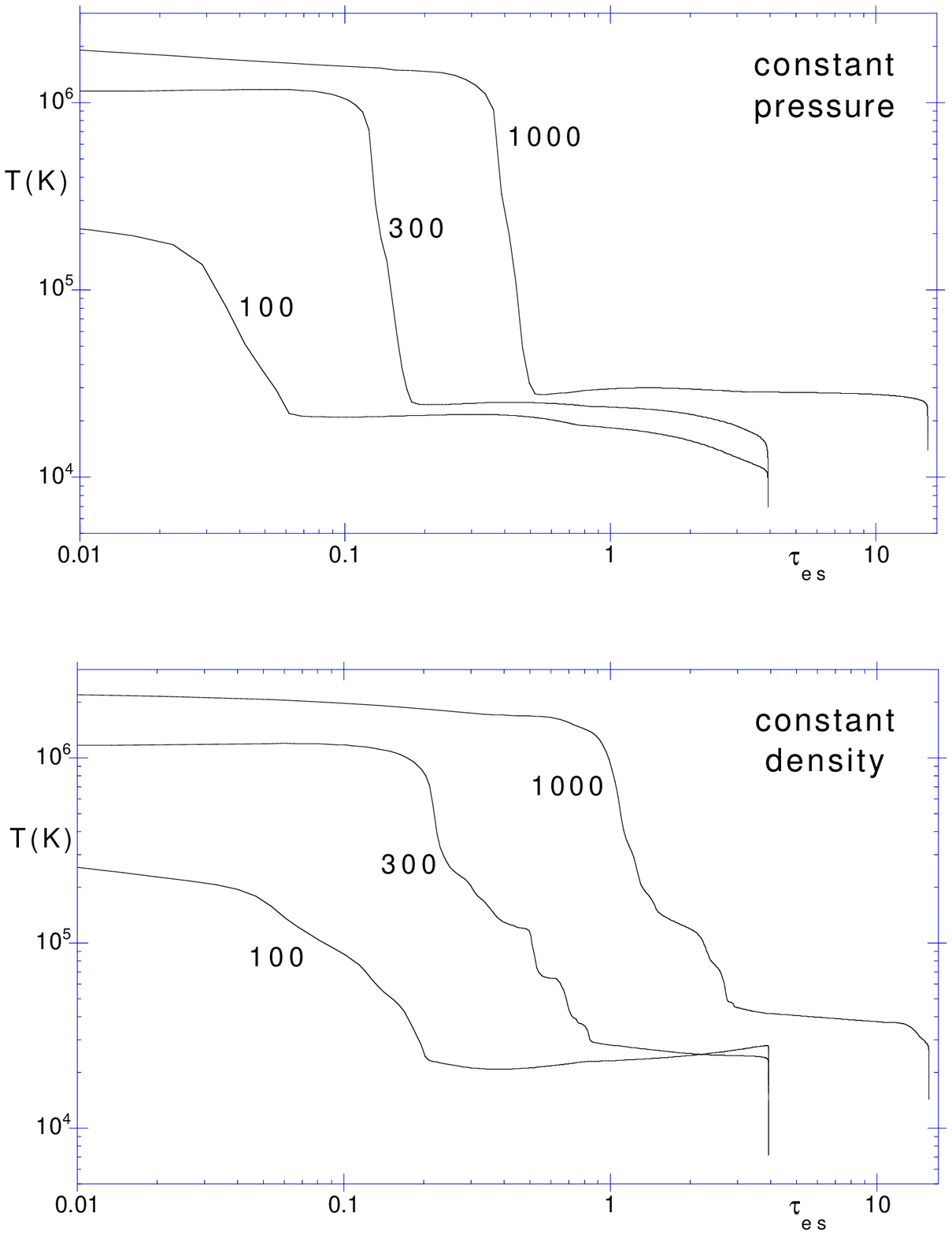}
\caption
{Temperature versus depth for different values of $\xi$ in erg
cm s$^{-1}$, for constant pressure and  constant density. The constant 
density slabs have a density of $10^{12}$ cm$^{-3}$, and the density at 
the surface of the slab with constant pressure is equal to the same value.
The other parameters are the same as in the reference model. }
\end{figure}


\begin{figure}
\plotone{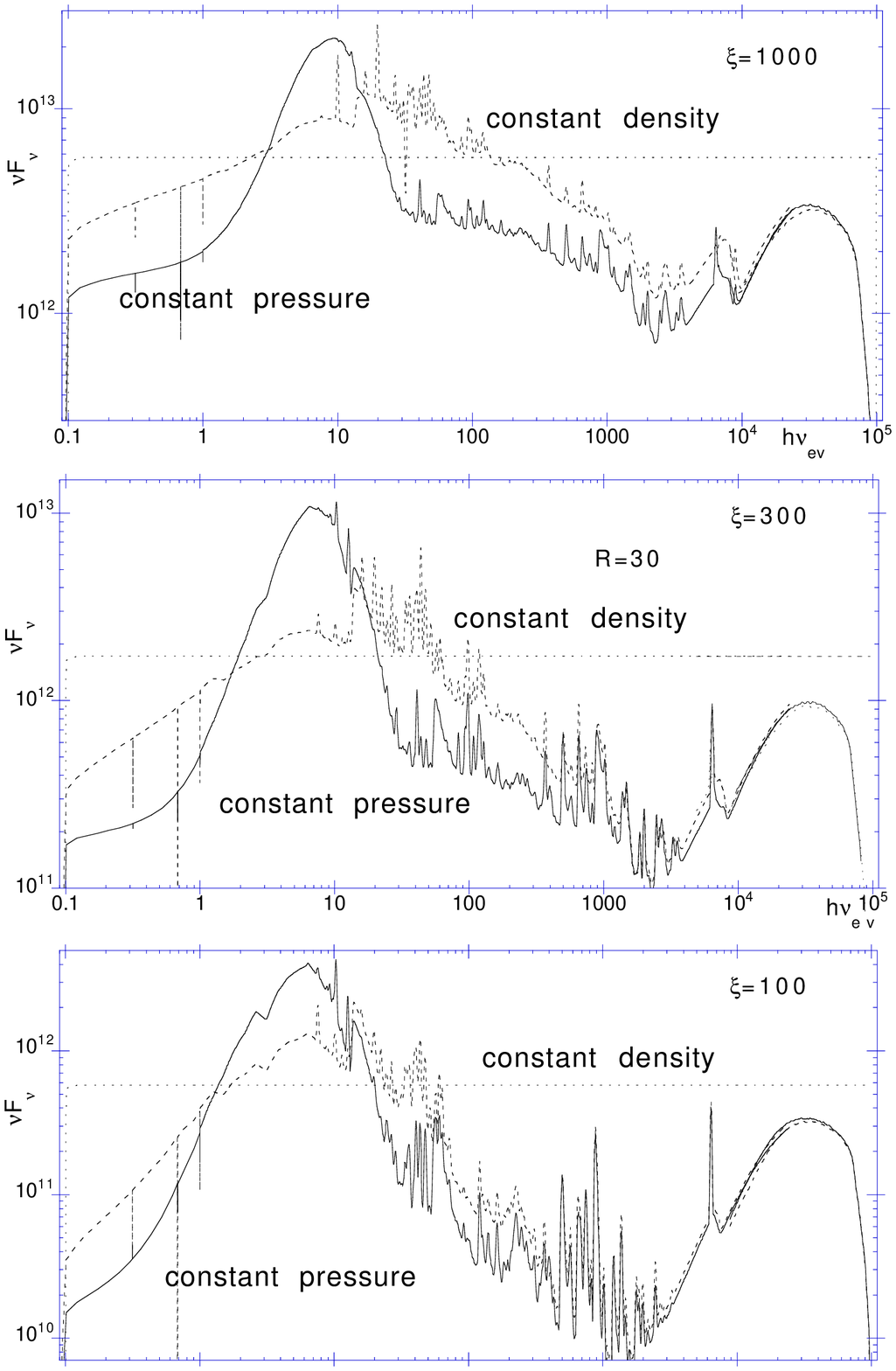}
\caption
{Reflected spectra computed with constant pressure (full line) and 
with constant density (dashed lines) for the models of the previous figure;
the spectra are displayed with a resolution of 30.}
\end{figure}

For constant pressure the spectrum displays a 
narrow UV bump, while the constant density the UV bump is 
more extended in frequency, as it corresponds to intermediate temperatures 
which do not exist in the constant pressure case (cf. Fig. 14). The difference between the
 two cases
 is particularly 
apparent for high 
illumination, as expected. It is interesting to note that the
Lyman edge is quite weak in the constant pressure case, since the UV spectrum 
is closer to a blackbody. This is 
a point that constant density models have difficulties to account for in AGN.


\begin{figure}
\plotone{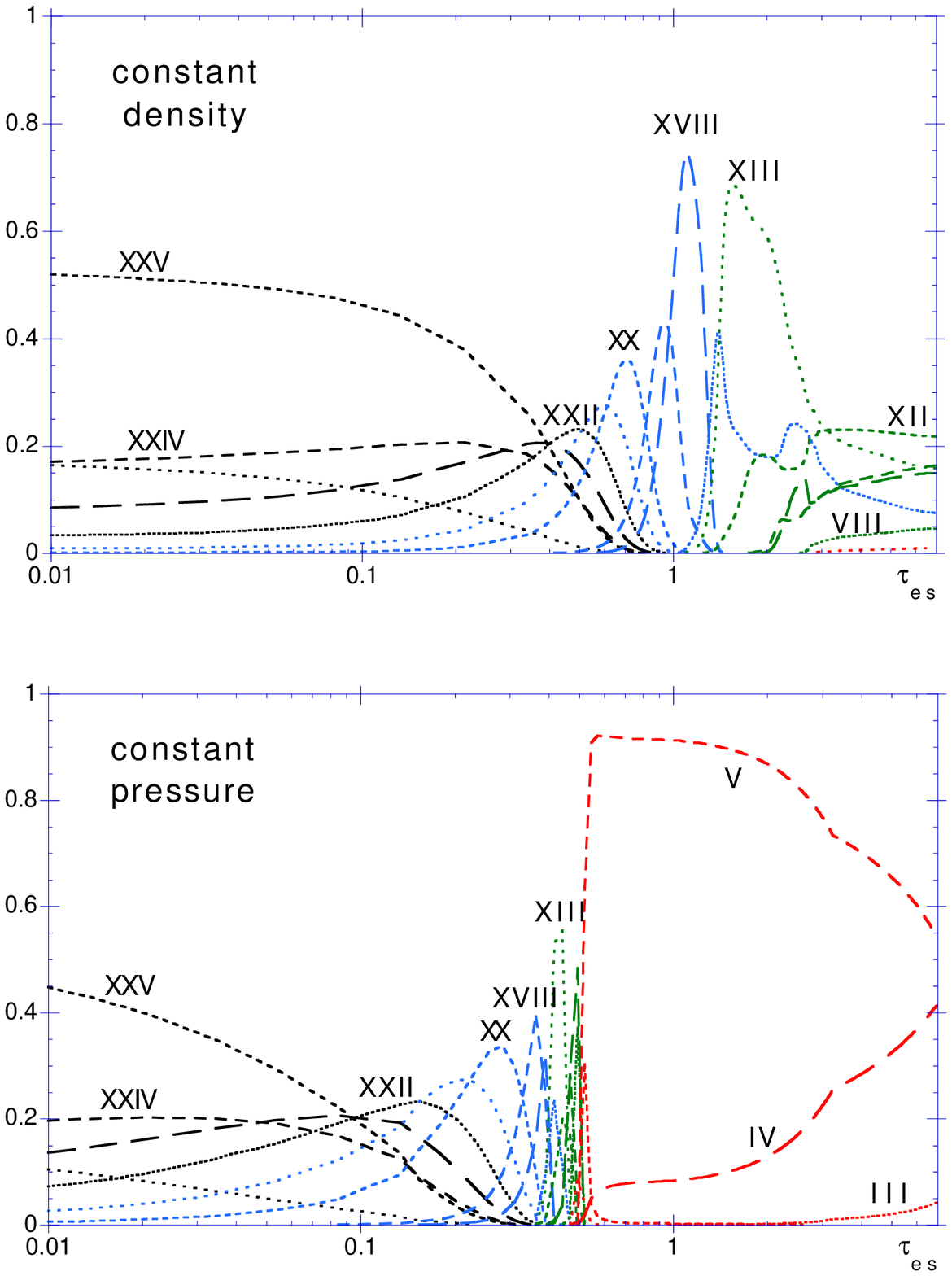}
\caption
{Fractional abundances of iron versus depth, in the case of 
constant pressure or constant density slabs, for the incident flux 
$8.10^{13}$ erg cm$^{-2}$ s$^{-1}$ for one of the model of the previous 
figure ($\xi=1000$).}
\end{figure}

The distribution of fractional ionic abundances is also completely modified. As an 
illustration, Fig. 15 displays the fractional abundances of iron versus depth. The 
most remarkable difference is the presence at 
$\tau_{\rm es}\sim 1$ of low 
ionization species, such as FeIV or FeV, which are completely absent in 
the constant density case.  
One notes also that the extension of highly ionized species is smaller than in 
the constant density case, as expected. For instance FeXIV is present up to  $\tau_{\rm es} =
 0.4$ for constant 
pressure, instead of
$1.2$ for constant density.
As a consequence, the Iron K lines produced are different from the constant density, 
as we shall see later.

\subsubsection{Hydrostatic equilibrium}:
\medskip

In order to
describe an irradiated accretion disk, we 
have introduced in collaboration with A. Rosanska and B. Czerny a variation of 
the density given by the hydrostatic equilibrium 
law. TITAN treats then only the layer at the surface of 
the disc where the diffusion approximation does not hold. In this case 
there is a flux 
incident on the back side of the slab, equal to the flux 
generated by viscous dissipation inside the disk.
Iterations are needed with the hydrostatic equilibrium equations (for more 
details see R\'o\.za\'nska et al. 2001).


\begin{figure}
\plotone{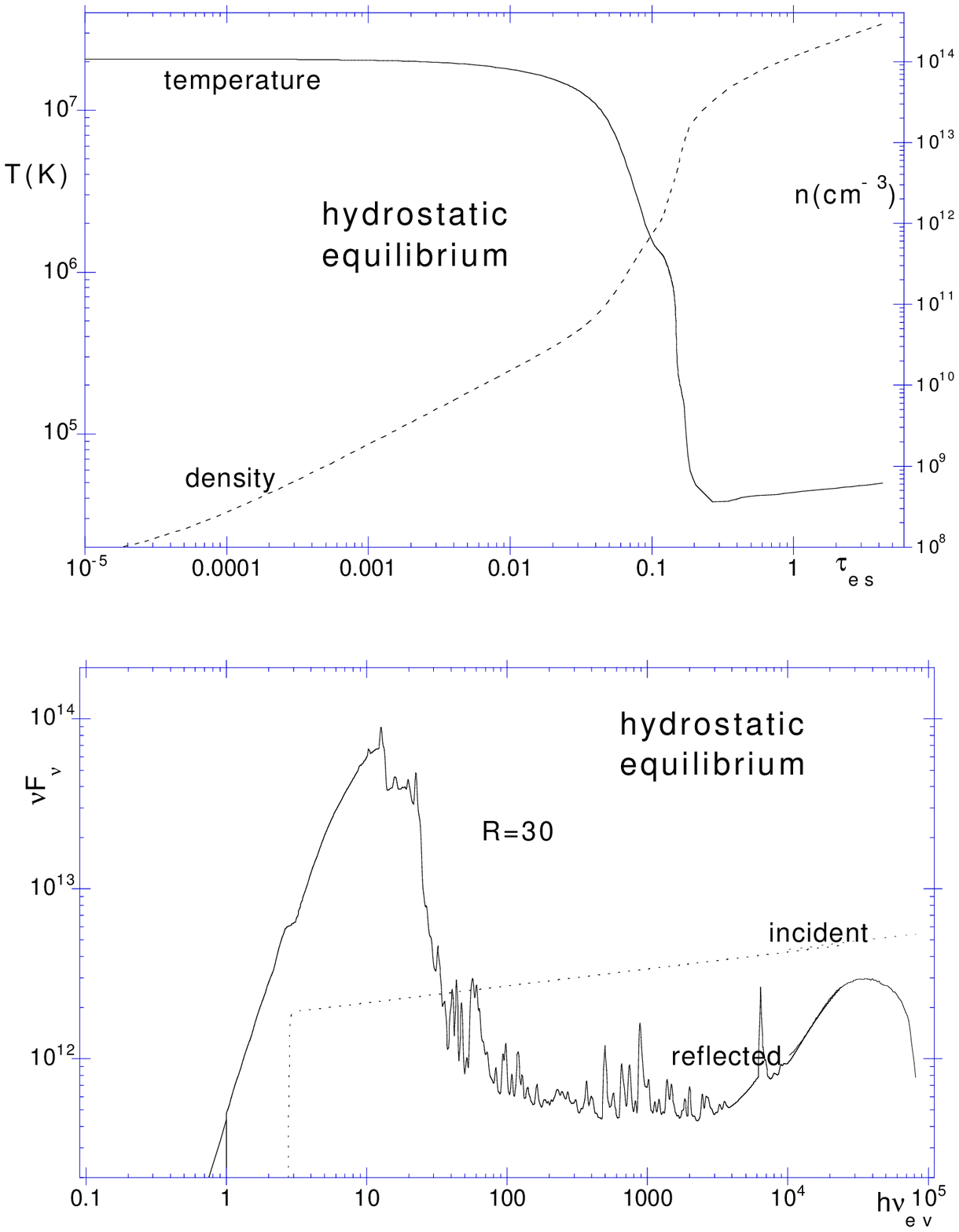}
\caption
{Temperature versus depth, and reflected spectrum, obtained 
for an illuminated accretion disk in hydrostatic equilibrium, as described 
in the text. From R\'o\.za\'nska et al., 2001.}
\end{figure}

Fig. 16 shows the results for the following case: $M_{\rm BH} = 10^8$ 
M$_{\odot} $, $\dot{m}=0.03$,  $R =10 R_{\rm 
sch}$, and an X-ray irradiating powerlaw spectrum, with a spectral 
index 0.9 in flux, and an X-ray flux $F_X=3.5\times 10^{13}$, i.e. equal to half
 the viscous flux.
Note that the temperature decreases abruptly like in the constant pressure 
case, and that the hot skin is very thin. The 
temperature reaches  that given by viscous dissipation at about 
$\tau_{\rm es} =0.4$. The reflected spectrum displays a prominent blue 
bump with no soft X-ray excess, and a strong absorption in the UV range due 
to low ionization species in the deep layers of the atmosphere. It is 
important to note that such a spectrum differs strongly from that of an 
irradiated disc in which the density is assumed to be constant (Ross \& 
Fabian 1993, and subsequent works, cf. R\'o\.za\'nska 
et al. 2001 for a detailed discussion). 
Our results are similar to those of Nayakshin, Kazanas \& Kallman (2000). 

\section{Coupling of TITAN and NOAR}

\subsection{Monte Carlo code NOAR}

In order to compute the effect of Compton scattering on the high 
energy part of the reprocessed radiation and to determine precisely the 
Compton heating-cooling rate in the energy balance equation,
one needs another numerical approach. In this aim Abrassart (1998) developed 
a Monte Carlo code, NOAR (cf. Dumont et al. 2000). The asset of such an approach is 
also that it enables to investigate an arbitrary geometry 
and to determine the angular dependence of the observed spectra.
Moreover, it allows to easily extract time variability information.

NOAR takes into account direct and inverse Compton scattering, 
according to the method proposed by Pozdniakov, Sobol \& Sunyaev (1983) and 
by Gorecki \& Wilczewski (1984). 
Given all the fractional abundances and the temperature in each layer provided by TITAN,
 it computes the 
absorption cross sections in each layer. Free-free absorption is 
taken into account, as well as recombination continua 
of hydrogen and helium like ions (the spectral distribution of the 
recombination continuum is determined by the local temperature).
The proper yields to account for the competition with the Auger effect
are used for the fluorescence lines. Fluorescence of Iron XVII to XXIII
is suppressed by resonant trapping (but not in TITAN). 
For the pseudo fluorescence of Iron XXV and XXVI, a simplifying 
prescription is adopted which includes the two most probable outcomes 
of a K shell photoionization, that is: direct recombination on ground 
level or L-K transition. Other line emission is not included.

One use of NOAR is to provide TITAN with the 
local Compton gains and losses in each layer. This is necessary, 
because Compton heating-cooling is dominated by energy losses of 
photons $>$ 20 keV for high values of $\xi$.
In Dumont et al. (2000) it was shown that given an incident spectrum
 the dependence of the Compton heating-cooling 
over $\xi$ ratio on the Thomson depth depends only slightly on
 $\xi$ except for high values of $\xi$.
The Compton heating-cooling rate 
obtained with NOAR is fitted analytically as a function 
of $z$, which is transferred to TITAN.
The spectra computed with TITAN are then completed. (cf. Fig. 17, and 
also Figs. 14 and 16).


\begin{figure}
\plotone{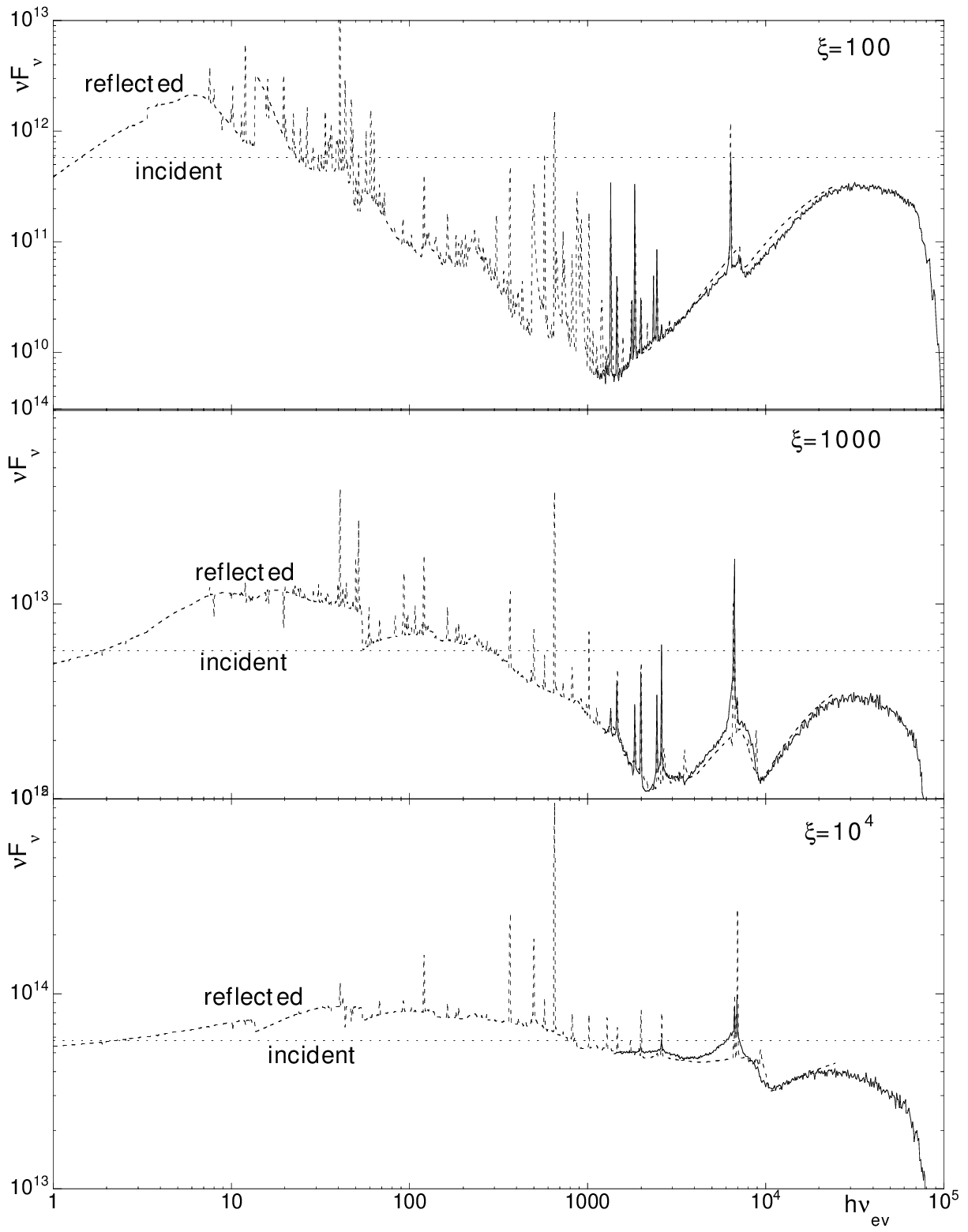}
\caption
{Reflected spectrum for different values of $\xi$ in erg cm s$^{-1}$,
the other parameters being the same as in the reference model; full lines:
results of NOAR; dashed lines: results of TITAN. 
The spectra are displayed with a resolution of 100.}
\label{fig-Gtitnoarx2-3-4}
\end{figure}

\subsection{Some results of TITAN and NOAR}

TITAN and NOAR lead to similar energy distributions of the reflected spectra 
in the 1-20 keV range. The reflected spectrum above 10 keV exhibits 
a ``Compton  hump'' depending on the high energy cut-off of the incident 
spectrum, which is here 100 keV, and on the   
ionization parameter.  The spectrum
 in the higher energy range is used 
for the energy balance. 

Comptonisation leads also to line broadening, and when it is important, to smearing of the 
ionization edges. Direct Compton downscatterings produce 
a red wing, while inverse Compton upscatterings produce a 
blue wing. The effect can be comparable to relativistic broadening near the 
black hole, and can account partly for the broad profiles of the iron line 
in AGN (Abrassart \& Dumont 1998). 

Fig. 18 displays (with a relatively 
high resolution, in order to show detailed features) the spectrum around 
7 keV in the constant density case, for different values of $\xi$, the 
other parameters being as in the reference model. 
The feature near 7 keV is made of a mixture of iron 
edges and of several iron lines:

- for $\xi$ equal to a few hundreds erg cm s$^{-1}$, the iron line is 
dominated by low ionization stages, smaller than iron XVII; the absorption 
edge is proeminent above 7 keV;

- for $\xi$ equal to $10^3$ erg cm s$^{-1}$, FeXXV 
and FeXXVI lines are apparent, modified by Compton scattering.
The lines are significantly broadened, the broadening is asymmetric, the 
profile is skewed towards the red. The absorption edge is still important, 
but extends from 8 to 10 keV, due both to Comptonization and to the 
influence of several edges;

- for $\xi$ equal to $10^4$ erg cm s$^{-1}$, the line has a large red 
Compton
wing and a weak blue wing, and it is weak. The absorption edge is completely
 erased.

Fig. 19 displays the iron line complex for the constant pressure case, to 
be
compared to the constant density case. Both a ``cold line" at 6.4 keV 
and FeXXV and XXVI lines  are present in the constant pressure case, while only 
highly 
ionized 
species are visible in the constant density case.
For constant pressure the cold layers are indeed closer from the surface 
than in the constant density case, so the 
photons emitted 
by these layers are able to leave the medium.


\begin{figure}
\plotone{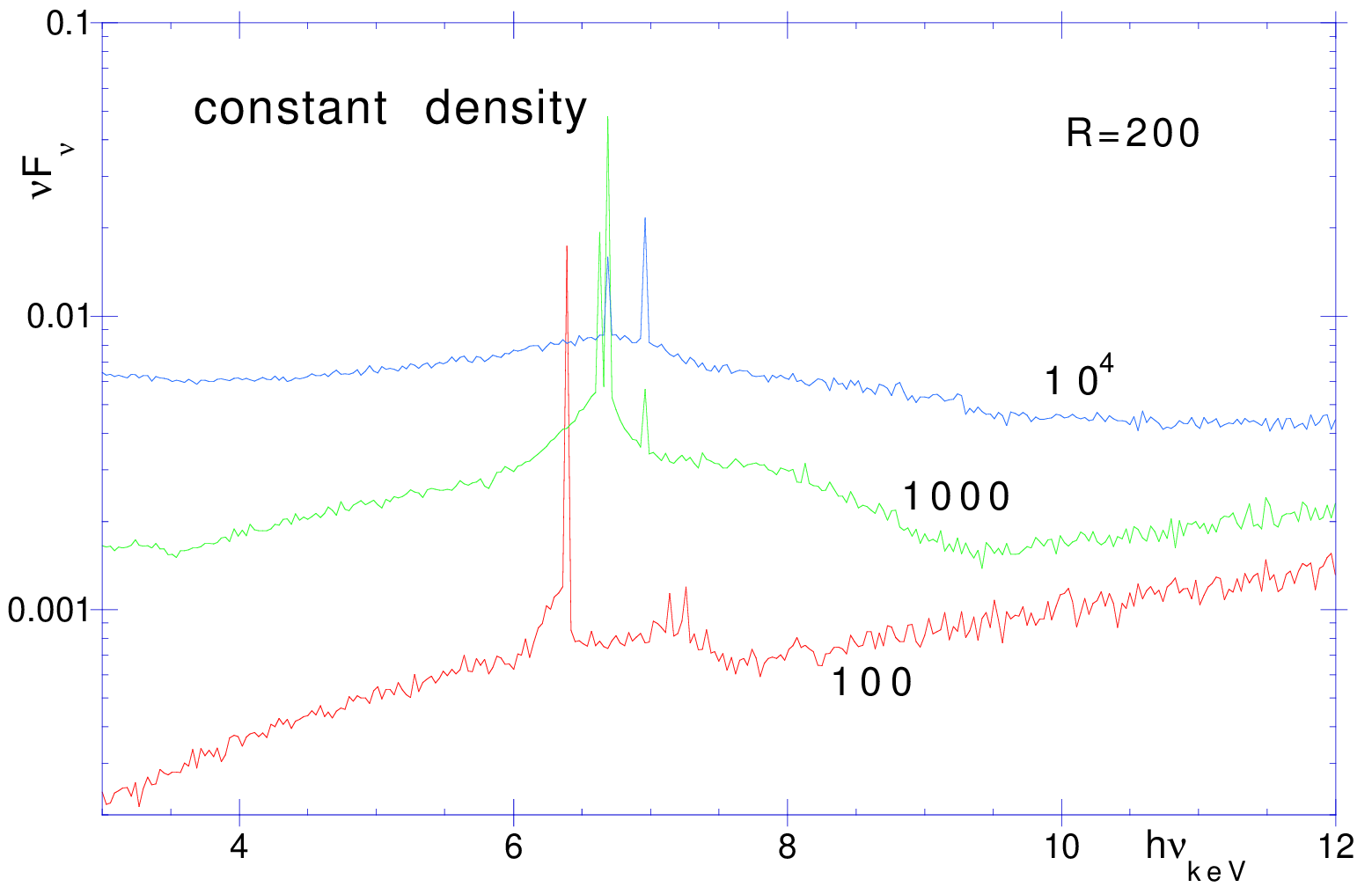}
\caption
{Reflected spectrum around 7 keV for different values of $\xi$ 
in erg cm s$^{-1}$,
the other parameters being the same as in the reference model;
The broadening by Comptonization is important for $\xi > 1000$ erg cm s$^{-1}$.
The spectra are displayed with a resolution of 200.}
\end{figure}


\begin{figure}
\plotone{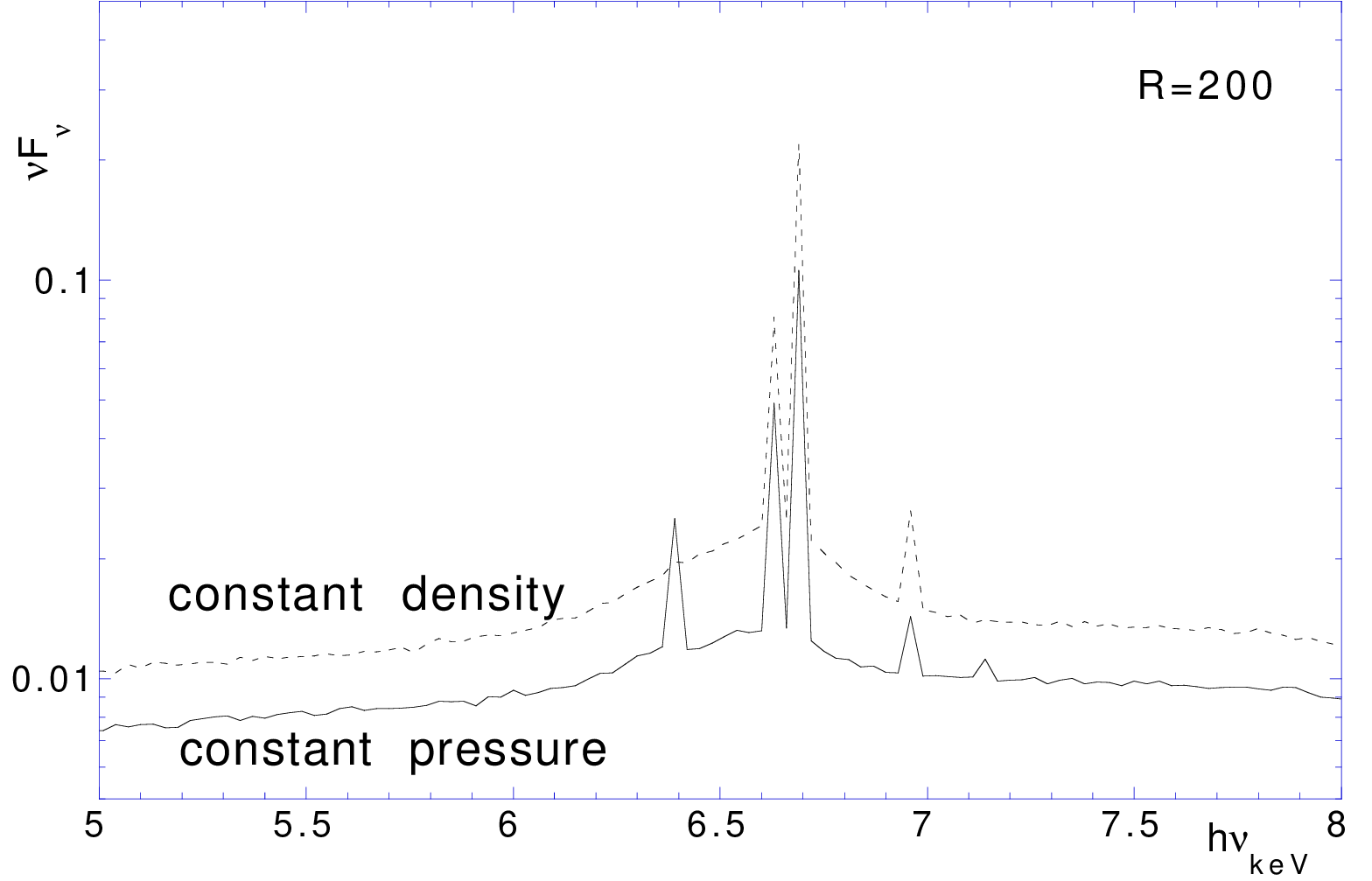}
\caption
{Reflected spectrum in the X-range in the case of
constant pressure or constant density slabs, for the incident flux
$8.10^{13}$ erg cm$^{-2}$ s$^{-1}$.
The spectra are displayed with a resolution of 200.}
\end{figure}
  
\section{Foreseen improvements}

The coupling of TITAN and NOAR
allows to compute the structure and the emission of hot Compton 
thick irradiated media, by solving consistently the energy balance of the 
medium. Moreover it allows also to take into account Compton broadening of the 
lines and of the edges in the X-ray range. 

In the previous description of the code (Dumont et al. 
2000), we had already
recalled the importance of 
the returning flux, which is often neglected, even for relatively 
low column densities, and we had compared the line transfer treatment 
for the lines with the escape probability approximation generally used in 
these problems. Here we have in addition stressed the influence of 
the pressure law, by showing the strong differences existing both in the structure 
and in the emitted spectrum, between a constant 
density medium, a constant pressure medium, or a medium in hydrostatic equilibrium. 
We have also shown that the atom
 models are of fundamental importance as they determine the 
 ionization state, the energy balance, the temperature, and in fine the 
 detailed spectral features which will be soon accessible through X-ray 
 spectroscopy. 

All along this paper we have mentioned what improvements we intend to 
bring to the code.

The Accelerated Lambda Iteration method is used now for the continuum. We 
will soon 
implement it 
for the lines, which are the most difficult to converge. 
It will in particular allow to treat partial redistribution in the lines.

 In collaboration with S. Coupe and M.C. Artru (Coupe et al. 2001)
we have begun to take into account subordinate lines for all ions by solving 
a complete multi-level atom for Li-like and He-like ions, after H-like ions 
which were already introduced in the previous version of the code. In the future 
we plan to implement other iso-electronic sequences, and to add to the 
present ones several other levels. In particular we will
try to get a better 
representation of the highest levels for computations close to LTE. 
We will take 
into account forbidden lines, which could be important for 
the cooling of the medium in some places. 
 We also intend to add to the line transfer an option 
allowing the UV and soft X-ray photons to escape from the medium through Compton
 scattering. 

Though there are still many improvements to perform, we presently
use 
these codes to model AGN spectra in the UV and X-ray range, in conditions where
 they
 are valid.

\end{document}